\documentclass[prd,floatfix,onecolumn,amsmath,amssymb]{revtex4-1}
\usepackage{graphicx,color,dcolumn,booktabs,bm}
\usepackage{subfigure}
\usepackage{amssymb}
\usepackage{longtable}
\usepackage{indentfirst}
\usepackage{epsfig,gensymb,siunitx}
\usepackage{feynmf}   
\usepackage{epstopdf}   
\usepackage{slashed}  
\usepackage{cases}
\usepackage[pdfpages]{xcolor}
\definecolor{maroon}{RGB}{139,25,150}
\usepackage{multirow}
\usepackage{float}
\usepackage{pgf}
\usepackage{graphicx,color,dcolumn,booktabs,bm}
\usepackage[colorlinks, citecolor=blue,anchorcolor=red,menucolor=red, linkcolor=red,filecolor=red,runcolor=red,urlcolor=blue,frenchlinks=red, urlcolor=blue]{hyperref}

\begin{document}

	\preprint{}
	
\title{\color{maroon}{Determination of the pion generalized parton distributions at zero skewness}}

	\author{The MMGPDs Collaboration:\\
Muhammad Goharipour$^{1,2}$}
\email{muhammad.goharipour@ipm.ir}
\thanks{Corresponding author}

\author{M.~H.~Amiri$^{3}$}
\email{mh.amiri79@ut.ac.ir}

\author{Fatemeh Irani$^{3}$}
\email{f.irani@ut.ac.ir}

\author{Hadi Hashamipour$^{4}$}
\email{hadi.hashamipour@lnf.infn.it}

\author{K.~Azizi$^{3,5,2}$}
\email{kazem.azizi@ut.ac.ir}

\affiliation{
$^{1}$School of Physics, Institute for Research in Fundamental Sciences (IPM), P.O. Box  19395-5531, Tehran, Iran\\
$^{2}$School of Particles and Accelerators, Institute for Research in Fundamental Sciences (IPM), P.O. Box 19395-5746, Tehran, Iran\\
$^{3}$Department of Physics, University of Tehran, North Karegar Avenue, Tehran 14395-547, Iran\\
$^{4}$Istituto Nazionale di Fisica Nucleare, Gruppo collegato di Cosenza, I-87036 Arcavacata di Rende, Cosenza, Italy\\
$^{5}$Department of Physics, Dogus University, Dudullu-\"{U}mraniye, 34775 Istanbul, T\"urkiye}	
\date{\today}

\begin{abstract}
We perform a global QCD analysis of the pion electromagnetic form factor (FF) data from pion electroproduction and elastic pion scattering to extract the valence pion generalized parton distributions (GPDs) at zero skewness. The analysis uses three different sets of pion parton distribution functions (PDFs), namely xFitter, JAM21, and MAP23, to construct the GPD ansatz. Through a $\chi^2$ minimization and a careful parametrization scan, we determine the profile function parameters and find that only two parameters are sufficient to describe the data. The extracted valence pion GPDs from different analyses have similar $x$-dependence, with minor differences at small momentum transfer. The resulting theoretical predictions for the pion electromagnetic FF and its squared magnitude show good agreement with experimental measurements. Among the three analyses, the one using the MAP23 PDFs provides the best overall fit and is adopted as the final GPD set. Our results offer a consistent determination of the valence pion GPDs, indicating a minor impact of the choice of pion PDFs. The present study provides a solid foundation for future investigations of pion structure, including its charge radius, tomography, and mechanical properties.

\end{abstract}

\maketitle

\section{Introduction}\label{sec:one} 

The pion plays an important role in our understanding of the strong interaction as described by quantum chromodynamics (QCD). It is the simplest quark-antiquark ($q\bar{q}$) bound state in the quark-parton model of hadrons. The pion mediates nucleon-nucleon forces and is one of the Nambu-Goldstone bosons associated with spontaneous chiral symmetry breaking in QCD, making it essential for explaining the origin of mass in hadronic matter (see, e.g., Refs.~\cite{Yndurain:2002ud,Horn:2016rip,Aguilar:2019teb,Ananthanarayan:2022wsl} for detailed discussions of pion physics). Despite being theoretically simpler than the proton, the internal structure of the pion remains less well understood, especially in comparison to that of the proton~\cite{Holt:2010vj}.

The concept of parton distribution functions (PDFs), which describe the probability densities of partons within a hadron as a function of the fraction of the hadron's momentum that they carry, is fundamental to high-energy particle physics, particularly in the description of scattering processes and collisions~\cite{Ethier:2020way,Lorce:2025aqp}. The extraction of PDFs is often formulated as a nonlinear regression problem, where the goal is to determine functional forms from experimental data (see, e.g., Refs.~\cite{H1:2015ubc,Accardi:2016qay,Alekhin:2017kpj,Hou:2019efy,Bailey:2020ooq,ATLAS:2021qnl,NNPDF:2021njg,PDF4LHCWorkingGroup:2022cjn} for unpolarized PDFs, and Refs.~\cite{deFlorian:2008mr,Leader:2010rb,Blumlein:2010rn,Nocera:2014gqa,Jimenez-Delgado:2014xza,Salajegheh:2018hfs,Adamiak:2023yhz,Bertone:2024taw,Borsa:2024mss,Cruz-Martinez:2025ahf} for polarized PDFs of the proton). These distributions are universal objects, meaning they can be applied across a wide range of processes involving parton dynamics. In this context, the determination of pion PDFs, either through global analyses of experimental data~\cite{Gluck:1991ey,Wijesooriya:2005ir,Aicher:2010cb,Han:2020vjp,Novikov:2020snp,Barry:2018ort,Barry:2021osv,Barry:2025wjx,Pasquini:2023aaf,Kotz:2023pbu,Kotz:2025lio,Good:2025nny} or via theoretical approaches~\cite{Gluck:1999xe,Abdel-Rehim:2015owa,Cui:2022bxn,Hutauruk:2023ccw,Lan:2019rba,Lan:2024ais,Chen:2024dhz,Maerovitz:2025txk,Kaur:2025gyr,Bopsin:2025vhz,Francis:2025rya,Wang:2025usl}, has attracted considerable interest in recent years. One can point to the efforts by the xFitter~\cite{Novikov:2020snp}, JAM~\cite{Barry:2018ort,Barry:2021osv,Barry:2025wjx}, and MAP~\cite{Pasquini:2023aaf} collaborations for the case of global analysis of pion PDFs.

Although PDFs provide valuable information about the internal structure of hadrons in the longitudinal (beam) direction, they cannot offer insight into the transverse spatial distribution of partons. In this regard, generalized parton distributions (GPDs) have emerged as a powerful framework for obtaining a more comprehensive picture of hadron structure~\cite{Ji:1998pc,Goeke:2001tz,Burkardt:2000za,Burkardt:2002hr,Polyakov:2002yz,Diehl:2003ny,Ji:2004gf,Belitsky:2005qn,Boffi:2007yc,Diehl:2015uka,Kumericki:2016ehc,Hashamipour:2022noy,Goharipour:2025kif}. GPDs correlate the transverse position and longitudinal momentum of partons, thus enabling a three-dimensional characterization of hadrons. They are also an essential ingredient in the theoretical description of hard exclusive and elastic scattering processes, as various moments of GPDs are related to different hadronic form factors (FFs)~\cite{Bernard:2001rs,Guidal:2004nd,Diehl:2013xca,Polyakov:2018zvc}. While significant progress has been made in the study of proton GPDs, the determination of pion GPDs has also attracted growing interest in recent years~\cite{Adhikari:2021jrh,Chavez:2021llq,Zhang:2021mtn,Zhang:2021shm,Raya:2021zrz,Cao:2021aci,Broniowski:2022iip,Lin:2023gxz,Xu:2023bwv,Ding:2024saz,Son:2024uet,Nematollahi:2024wrj,Puhan:2025ibn,Nasibova:2025wnw,Zhang:2025xtn}.

The electromagnetic FF of the pion is a fundamental observable that encodes essential information about its internal structure and spatial charge distribution. The pion is the simplest mesonic system composed of a quark and an antiquark. So, it offers a clean environment to study the dynamics of the strong interaction in both perturbative and nonperturbative regimes of QCD. Study of the  pion FF provides a critical testing ground for a variety of theoretical approaches~\cite{Lepage:1980fj,Bakulev:2004cu,Chen:2023byr,Wang:2025irh,Boyle:2008yd,Wang:2020nbf,Ding:2024lfj,Gao:2021xsm,Chang:2013nia,Jia:2024dfl,Brodsky:2007hb,Simula:2023ujs,Nesterenko:1982gc,Ayala:2025wuu,Xu:2023izo,Yao:2024drm,Leao:2024agy,Kirk:2024oyl,Miramontes:2024fgo,Ji:2024iak,Puhan:2025pfs}, including perturbative QCD~\cite{Lepage:1980fj,Bakulev:2004cu,Chen:2023byr,Wang:2025irh}, lattice QCD~\cite{Boyle:2008yd,Wang:2020nbf,Ding:2024lfj,Gao:2021xsm}, the Dyson--Schwinger equation (DSE)~\cite{Chang:2013nia,Jia:2024dfl}, light-front holography (LFH)~\cite{Brodsky:2007hb}, the dispersive matrix approach~\cite{Simula:2023ujs}, the light-cone quark model (LCQM)~\cite{Puhan:2025pfs}, and QCD sum rules~\cite{Nesterenko:1982gc,Ayala:2025wuu}. On the experimental side, the spacelike pion FF has been measured via pion electroproduction processes such as $ep \rightarrow e'\pi^+n$~\cite{Brown:1973wr,Ackermann:1977rp,Bebek:1977pe,Brauel:1979zk,JeffersonLabFpi:2000nlc,JeffersonLab:2008jve} and elastic pion scattering~\cite{Adylov:1977kj,Dally:1981ur,Dally:1982zk,NA7:1986vav}. The timelike region has also been explored in $e^+e^- \rightarrow \pi^+\pi^-$ annihilation~\cite{CMD-2:2001ski,Achasov:2006vp}. A very important point in this regard is that the pion FF is closely related to the first moment of the pion GPD, making it a vital input and constraint in the modeling and extraction of the pion's GPDs. Thus, understanding the pion FF not only sheds light on the nature of confinement and hadronization in QCD but also provides key input for the determination of the pion's GPDs. 
In addition to the electromagnetic form factor, the gravitational FFs (GFFs) of the pion, which are related to moments of the energy-momentum tensor, have also attracted significant interest in recent years~\cite{Hatta:2025ryj,Tanaka:2025znc,Choi:2025rto,Voronin:2025sbs,Fujii:2025tpk}.

Following the MMGPDs studies on the extraction of proton GPDs in the zero-skewness limit~\cite{Goharipour:2024atx,Goharipour:2024mbk}, in the present work, we aim to determine the pion GPDs through a global analysis of all available pion FF data from both pion electroproduction and elastic pion scattering. This is the first application of the MMGPDs framework to the pion, extending its utility beyond the proton sector. To this end, we adopt the same functional ansatz used in the proton analysis and investigate its sensitivity to the choice of input pion PDFs. In Sec.~\ref{sec:two}, we introduce the pion PDF sets used in our analysis, highlighting their key features and differences. In Sec.~\ref{sec:three}, we introduce the phenomenological framework used to extract the pion GPDs from the FF data.  Section~\ref{sec:four} is devoted to the experimental data selection and treatment. A detailed discussion of the extracted GPDs and a comparison with the input data is presented in section Sec.\ref{sec:five}. Finally, a summary and our conclusions are given in Sec.~\ref{sec:six}.

\section{Pion PDFs}\label{sec:two}

Pion PDFs play an important role in our understanding of hadron structure and QCD dynamics. As the lightest meson and the simplest quark-antiquark bound state, the pion provides a unique framework for  testing theoretical models of parton dynamics, confinement, and chiral symmetry breaking. In particular, pion PDFs are essential inputs in the theoretical description of high-energy processes involving pions, such as Drell-Yan (DY) production~\cite{Conway:1989fs,NA10:1985ibr,Stirling:1993gc}, leading neutron (LN) electroproduction~\cite{Feruglio:2002af,H1:2010hym}, and prompt photon production~\cite{WA70:1987bai}. They can also play a crucial role in the extraction of pion GPDs, where the forward limit of the GPD is directly related to the PDF. Unlike proton PDFs, which are constrained by a wide range of precise experimental data, pion PDFs are less well known due to the absence of stable pion targets. However, several recent efforts have provided improved extractions of pion PDFs through both data analysis~\cite{Han:2020vjp,Novikov:2020snp,Barry:2018ort,Barry:2021osv,Pasquini:2023aaf,Kotz:2023pbu,Kotz:2025lio,Good:2025nny,Barry:2025wjx} and theoretical calculations~\cite{Cui:2022bxn,Hutauruk:2023ccw,Lan:2019rba,Lan:2024ais,Chen:2024dhz,Maerovitz:2025txk,Kaur:2025gyr,Bopsin:2025vhz,Francis:2025rya,Wang:2025usl}.
The recent determinations of the pion PDFs have adopted distinct frameworks, methodologies, and theoretical inputs, reflecting complementary approaches to constraining the pion's partonic structure. In this section, we outline the defining features of the xFitter~\cite{Novikov:2020snp}, JAM21~\cite{Barry:2021osv}, and MAP23~\cite{Pasquini:2023aaf} pion PDF sets, highlighting their key features and differences.

The xFitter Collaboration has provided an open-source pion PDF fit based on a combination of pion-induced DY and prompt-photon production data for the first time.~\cite{Novikov:2020snp}. In this phenomenological approach, flexible functional forms are used to parametrize the PDFs at an input scale, and DGLAP evolution is applied to evolve them to experimental scales.
An important property of the xFitter analysis is its comprehensive treatment of uncertainties which incorporates both experimental and theoretical uncertainties. The systematic uncertainties include also the scale variations and $\alpha_s$ dependence. Prompt-photon data included in the xFitter analysis play a crucial role considering the fact that they offer direct sensitivity to the gluon PDF. However, their impact is reduced by statistical limitations and theoretical uncertainties in the photon sector.

The JAM Collaboration has performed a global QCD analysis using pion-induced DY and LN electroproduction data~\cite{Barry:2021osv} (we call it JAM21 in this paper). As a distinguishing feature of this study, one can point to the systematic implementation of threshold resummation at next-to-leading logarithmic accuracy. The aim was to investigate the sensitivity of high-$x$ observables to soft-gluon emissions near the partonic threshold, where $x$ denotes the longitudinal momentum fraction of the pion carried by a constituent parton. JAM21 explored multiple resummation schemes, including Mellin-Fourier and double-Mellin techniques, to test the universality of resummation corrections to the large-$x$ behavior of the valence distribution. The analysis demonstrated that resummation can significantly affect the final results, particularly at larger values of $x$.
It should be noted that JAM21 considers PDFs as parametric functions (just like xFitter) constrained by data, and uses Bayesian methods to explore uncertainties. Sea and gluon distributions are allowed to vary and are constrained by the combined DY and LN datasets. This provides a more data-driven evaluation of the pion's partonic content at various scales.

The MAP Collaboration (referred to as MAP23 in this paper) has utilized a light-front constituent quark model approach rooted in an explicit Fock-space expansion of the pion state up to four-parton configurations, including $q\bar{q}$, $q\bar{q}g$, $q\bar{q}gg$, and $q\bar{q}q\bar{q}$ sectors~\cite{Pasquini:2023aaf}. The model directly encodes valence, sea, and gluon contributions into the structure of the light-front wave functions (LFWFs).
An important feature of the MAP23 analysis is the factorized parameterization of the longitudinal and transverse momentum components of the LFWFs. In this approach, the information on the longitudinal profile comes from the pion distribution amplitudes and the transverse profile is constrained by fits to the electromagnetic FF.
This model offers a complementary, theory-driven perspective on pion PDFs which link them explicitly to hadronic wave functions and form-factor phenomenology. 

\begin{figure}[!htb]
    \centering
\includegraphics[scale=0.5]{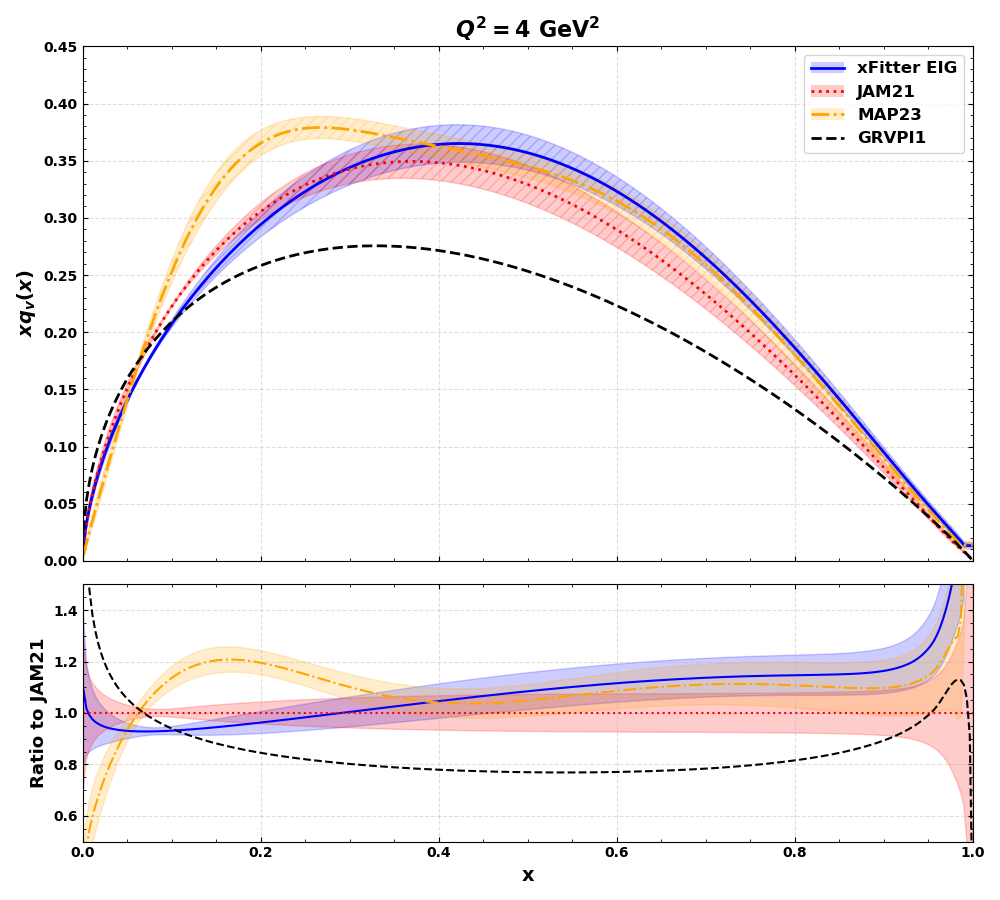}   
\caption{Comparison of the valence pion PDF $q_v(x)$ at $Q^2 = 4~\mathrm{GeV}^2$ from the xFitter~\cite{Novikov:2020snp}, JAM21~\cite{Barry:2021osv}, MAP23~\cite{Pasquini:2023aaf}, and GRVPI1~\cite{Gluck:1991ey} determinations. For xFitter, the EIG grid is used. The lower panel displays the ratio of each PDF to the JAM21 result.}
\label{fig:PDFs}
\end{figure}

Figure~\ref{fig:PDFs} compares the valence pion PDF, $ xq_v(x,Q^2)$, at $Q^2 = 4~\mathrm{GeV}^2$ from the xFitter~\cite{Novikov:2020snp}, JAM21~\cite{Barry:2021osv}, and MAP23~\cite{Pasquini:2023aaf} analyses, along with the older GRV determination~\cite{Gluck:1991ey} (denoted GRVPI1). Assuming charge symmetry for the valence quark PDF, we have $ q_v \equiv \bar{u}_v^{\pi^-} = \bar{u}^{\pi^-} - u^{\pi^-} = d_v^{\pi^-} $.
For xFitter, the EIG grid was employed in the comparison. To emphasize the differences, the lower panel of Fig.~\ref{fig:PDFs} shows the ratios of each result to JAM21. Distinct deviations appear mainly in the small- and large-$x$ regions, while xFitter and JAM21 agree more closely within their respective uncertainty bands. Such differences can have a non-negligible impact on phenomenological applications involving pion PDFs, including the present analysis of pion GPDs. In this work, we explicitly examine this sensitivity by repeating our extraction with different pion PDF inputs to assess the degree of PDF dependence in the final results.
%

\section{Phenomenological Framework}\label{sec:three}

As mentioned in the Introduction, in the present study, we are going to determine the pion GPDs through a global analysis of all available pion electromagnetic FF data from both pion electroproduction and elastic pion scattering (see the next section). In this section, we present the phenomenological framework used in our analysis. Note that the pion electromagnetic FF is evaluated in the spacelike region, $ t<0 $, where $t$ is the Mandelstam variable representing the squared four-momentum transfer. This kinematic domain is where GPDs are defined and can be interpreted in terms of PDFs within the pion. The $ t $-dependence of GPDs encodes information about the spatial distribution of quarks inside the pion.

The contribution of a given quark flavor $q$ to the pion electromagnetic FF, denoted by $F_\pi^q(t)$, can be expressed in terms of the valence unpolarized pion GPD $H_v^q(x,t,Q^2)$ as~\cite{Diehl:2013xca,Goharipour:2024atx}
\begin{equation}
F_\pi^q(t) = \int_{0}^1 dx \, H_v^q(x,t,Q^2) \, .
\label{Eq1}
\end{equation}
Here, $x$ is the longitudinal momentum fraction carried by the quark, and $Q^2$ is the scale at which the internal structure of pion is probed. In principle, GPDs also depend on the skewness parameter $\xi$, the longitudinal momentum transfer. Equation~\eqref{Eq1} holds for any value of $\xi$, but the $\xi$-dependence drops out upon integration over $x$ due to the polynomiality property~\cite{Ji:1998pc,Diehl:2003ny}. This kind of GPDs which is related to the FFs is commonly referred to as the zero-skewness GPDs~\cite{Hashamipour:2022noy}.
Considering the quark content of the charged pions ($u\bar{d}$ for $\pi^+$ and $d\bar{u}$ for $\pi^-$) and the corresponding quark electric charges $e_q$, the total pion FF can be written as
\begin{equation}
F_\pi(t) = \sum_q e_q \, F_\pi^q(t) \, .
\label{Eq2}
\end{equation}

Equation~\eqref{Eq1} expresses the sum rule connecting the pion electromagnetic FF to the first moment of the GPD at fixed scale $ Q^2 $. This relation ensures that the GPDs are properly normalized and that the charge conservation condition 
$ F_\pi(0)=1 $ is automatically satisfied when the valence distributions are used. In the zero-skewness limit ($ \xi=0 $), which corresponds to the case where no longitudinal momentum is transferred between the initial and final pion states, the GPDs reduce to the form relevant for describing elastic scattering processes. 

To extract the pion GPDs from FF data, we adopt a functional form inspired by previous MMGPD studies on the proton~\cite{Goharipour:2024atx,Goharipour:2024mbk}:
\begin{equation}
H_v^q(x,t,Q^2) = q_v(x,Q^2) \, \exp\!\left[ t \, f_v^q(x) \right] ,
\label{Eq3}
\end{equation}
where $q_v(x,Q^2)$ is the valence PDF of flavor $q$ and $f_v^q(x)$ is a profile function encoding the $t$-dependence of the GPD. 
The factorized ansatz in Eq.~\eqref{Eq3} is widely used in phenomenological studies because it cleanly separates the longitudinal-momentum structure, encoded in the PDF $ q_v(x,Q^2) $, from the transverse spatial structure, governed by the exponential profile function $f_v^q(x)$. The exponential form ensures positivity and smoothly reproduces known Regge behavior at the respective kinematic limits. This approach has been successfully applied to nucleon and pion GPDs in previous analyses and provides a practical framework for fitting experimental data.
In this work, we take $q_v(x,Q^2)$ from the xFitter~\cite{Novikov:2020snp}, JAM21~\cite{Barry:2021osv}, and MAP23~\cite{Pasquini:2023aaf} analyses at $Q^2=4~\mathrm{GeV}^2$, as provided through the \texttt{LHAPDF} package~\cite{Buckley:2014ana}. This allows us to test the sensitivity of our results to the choice of pion PDFs and identify the set that best describes the data.

Following Refs.~\cite{Diehl:2013xca,Goharipour:2024atx,Goharipour:2024mbk}, we parametrize the profile function as
\begin{equation}
f_v^q(x) = \alpha' (1-x)^3 \ln\frac{1}{x} + B (1-x)^3 + A \, x (1-x)^2 \, ,
\label{Eq4}
\end{equation}
where $\alpha'$, $A$, and $B$ are free parameters to be determined separately for each quark flavor. This functional form ensures Regge-like behavior at small $x$ through the logarithmic term and a power-law falloff at large $x$.
To be more precise, the logarithmic term $ \ln(1/x) $ controls the small $ x $ rise of the GPD, consistent with the Regge trajectory slope $ \alpha' $, while the power terms in $ (1-x) $ govern the falloff at large $ x $, corresponding to the dominance of configurations where the struck quark carries most of the pion's momentum. Therefore, the parameters $\alpha'$, $A$, and $B$  have direct physical meaning: $\alpha'$ determines the effective Regge slope, while $ A $ and $ B $ adjust the transition between the nonperturbative and perturbative regimes.

The optimal values of these parameters are obtained by minimizing the $\chi^2$ function:
\begin{equation}
\chi^2 = \sum_{i=1}^n \left( \frac{{\cal E}_i - {\cal T}_i}{\delta{\cal E}_i} \right)^2 ,
\label{Eq5}
\end{equation}
where ${\cal E}_i$ is the $i$th experimental data point, ${\cal T}_i$ is the corresponding theoretical prediction, and $\delta{\cal E}_i$ is the total experimental uncertainty which is obtained by adding statistical and systematic errors in quadrature. The sum runs over all data points from the datasets discussed in Sec.~\ref{sec:four}.
In practice, we performed a systematic parameter-scan procedure~\cite{H1:2009pze} to identify the most stable minimum. This reduces the risk of local minima and ensures that the extracted parameters correspond to a genuine global optimum. The fit quality for each dataset and the resulting parameter uncertainties are reported in Sec.~\ref{sec:five}.

The minimization is carried out using the Python interface of the CERN \texttt{MINUIT} library~\cite{James:1975dr}, \texttt{iMINUIT}~\cite{iminuit}. The uncertainties on the extracted GPDs are estimated using the standard Hessian method~\cite{Pumplin:2001ct} as implemented in \texttt{iMINUIT}, with $\Delta\chi^2 = 1$ corresponding to the $68\%$ confidence level.

%

\section{Data Selection}\label{sec:four}

Our analysis relies on a comprehensive set of experimental measurements of the charged pion electromagnetic FF, $F_\pi(t)$, as well as its absolute square, $|F_\pi(t)|^2$, obtained from two main sources: pion electroproduction and elastic pion scattering. Pion electroproduction involves measurements of the pion FF extracted from the longitudinal cross section of the exclusive process $e p \to e' \pi^+ n$, where a leading neutron is produced in the final state. Elastic pion scattering provides direct measurements of $|F_\pi(t)|^2$ from electron-pion ($e$-$\pi$) scattering experiments.
In the present analysis, the momentum transfer $t$ lies in the spacelike region ($t < 0$), consistent with the kinematics of pion electroproduction and elastic scattering experiments from which the data are taken.  

The full list of datasets used in this study is summarized in Table~\ref{tab:datasets}, together with the corresponding kinematic ranges and the number of data points, $N_{\mathrm{pts}}$. The electroproduction category includes 40 data points covering the range $0.18 < |t| < 9.77~\mathrm{GeV}^2$. For the Brauel 1979 dataset~\cite{Brauel:1979zk}, we use the reanalyzed value reported in Volmer 2001~\cite{JeffersonLabFpi:2000nlc}. The elastic scattering category is dominated by classic low-$|t|$ data, including 101 data points covering the range $0.0138 < |t| < 0.092~\mathrm{GeV}^2$.  These measurements provide crucial constraints on the slope of the form factor at the origin and,
hence, on the pion charge radius. Since these data are free from the model assumptions inherent in electroproduction extractions, they play an important role in our global fit.

The combined dataset, including 141 data points, spans a broad kinematic range in $t$, from the near-forward region $|t|= 0.0138~\mathrm{GeV}^2$ (from elastic scattering) up to $|t|= 9.77~\mathrm{GeV}^2$ (from electroproduction). This coverage ensures sensitivity to both the long-distance and short-distance structure of the pion. Before fitting, we examined all datasets for potential inconsistencies by comparing overlapping measurements in the same $t$ region. We didn't find statistically significant tensions.
\begin{table}[!htb]
    \centering
    \caption{Summary of the experimental datasets used in this analysis. The kinematic coverage is given in terms of the squared momentum transfer $t$, and $N_{\mathrm{pts}}$ denotes the number of data points included from each source.}
    \label{tab:datasets}
    \begin{tabular}{l@{\hskip 0.7cm}c@{\hskip 0.7cm}c@{\hskip 0.7cm}c}
        \hline
        \textbf{Experiment} & \textbf{Process} & \textbf{$-t$ range (GeV$^2$)} & \textbf{$N_{\mathrm{pts}}$} \\
        \hline
        Brown 1973~\cite{Brown:1973wr} & $e p \to e' \pi^+ n$ & $0.176-1.188$ & $5$ \\        
        Ackermann 1978~\cite{Ackermann:1977rp} & $e p \to e' \pi^+ n$ & $0.35$ & $1$ \\
        Bebek 1978~\cite{Bebek:1977pe} & $e p \to e' \pi^+ n$ & $0.18-9.77$ & $21$ \\
        Brauel 1979~\cite{Brauel:1979zk} & $e p \to e' \pi^+ n$ & $0.7$ & $1$ \\        
        Volmer 2001~\cite{JeffersonLabFpi:2000nlc} & $e p \to e' \pi^+ n$ & $0.6-1.6$ & $4$ \\         
        Huber 2008~\cite{JeffersonLab:2008jve} & $e p \to e' \pi^+ n$ & $0.6-2.45$ & $8$ \\  
        Adylov 1977~\cite{Adylov:1977kj} & $e$-$\pi$ scattering & $0.0138-0.0353$ & $22$ \\ 
        Dally 1981~\cite{Dally:1981ur} & $e$-$\pi$ scattering & $0.0317-0.0705$ & $20$ \\     
        Dally 1982~\cite{Dally:1982zk} & $e$-$\pi$ scattering & $0.039-0.092$ & $14$ \\               
        Amendolia 1986~\cite{NA7:1986vav} & $e$-$\pi$ scattering & $0.015-0.253$ & $45$ \\ 
        \hline
        \textbf{Total} & -- & $0.0138-9.77$ & 141 \\
        \hline
    \end{tabular}
\end{table}

%

\section{Results}\label{sec:five}

In this section, we present the results of our global QCD analyses of the pion electromagnetic FF data. The fits have been performed using the three different pion PDF sets introduced in Sec.~\ref{sec:two}, namely xFitter~\cite{Novikov:2020snp} (EIG grid), JAM21~\cite{Barry:2021osv}, and MAP23~\cite{Pasquini:2023aaf}, to construct the GPD ansatz of Eq.~\eqref{Eq3}. In addition to determining the pion GPDs, we also investigate the sensitivity of the results to the choice of input PDFs. The GPD sets extracted from these three analyses will be referred to as Set~1, Set~2, and Set~3, respectively. The pion PDFs are taken at $ Q^2 = 4 \, \text{GeV}^2 $ using the \texttt{LHAPDF} package~\cite{Buckley:2014ana}, in analogy with previous proton GPD analyses~\cite{Hashamipour:2022noy,Goharipour:2024atx}. For each analysis, the optimum values of the free parameters in the profile function $ f_v^q(x) $ [see Eq.~\eqref{Eq4}] are obtained by minimizing the global $\chi^2$ function defined in Eq.~\eqref{Eq5}.

Following the standard assumption adopted in pion PDF analyses, namely $ \bar{u}_v^{\pi^-} = d_v^{\pi^-} $, we also assume that the profile functions of the constituent quarks are identical, so that a single profile function $ f_v(x) $ needs to be determined. This reduces the number of free parameters from six to three. Furthermore, based on the parametrization scan procedure used in previous analyses~\cite{Hashamipour:2022noy,Goharipour:2024atx}, originally proposed in the QCD analysis of PDFs by the H1 and ZEUS Collaborations~\cite{H1:2009pze}, we find that it is sufficient to vary only the parameters $\alpha'$ and $A$ in order to achieve a satisfactory description of the data. In fact, allowing parameter $B$ to vary does not lead to a significant reduction in the total $\chi^2$. Physically, $\alpha'$ and $A$ control the $t$-dependence of the GPDs at small and large $-t$, respectively, while $B$ mainly affects the intermediate $-t$ region. Our investigations indicate that a full description of the data can be achieved by considering only $\alpha'$ and $A$.

Table~\ref{tab:fitparams} summarizes the best-fit values of the profile function parameters, $\alpha'$ and $A$, for each analysis. The extracted parameters exhibit a mild dependence on the choice of pion PDF set, while the overall trends remain consistent across the three fits. In particular, we observe an inverse correlation: as the value of $\alpha'$ increases, the corresponding value of $A$ decreases. The quoted uncertainties represent the $1\sigma$ confidence intervals, obtained using the Hessian method described in Sec.~\ref{sec:three}.
\begin{table}[!htb]
    \centering
    \caption{Best-fit values of the GPD profile function parameters with their uncertainties, obtained from analyses using different pion PDF sets. Set~1, Set~2, and Set~3 correspond to the xFitter~\cite{Novikov:2020snp}, JAM21~\cite{Barry:2021osv}, and MAP23~\cite{Pasquini:2023aaf} pion PDFs, respectively.}
    \label{tab:fitparams}
    \begin{tabular}{l@{\hskip 1.5cm}c@{\hskip 1.5cm}c}
        \hline
        \textbf{GPD set} & $\boldsymbol{\alpha'}$ & $\boldsymbol{A}$  \\
        \hline
        Set~1 & $ 1.008 \pm 0.024 $ & $ 2.929 \pm 0.146 $  \\
        Set~2 & $ 1.123 \pm 0.023 $ & $ 2.724 \pm 0.146 $  \\
        Set~3 & $ 1.196 \pm 0.024 $ & $ 2.384 \pm 0.148 $  \\
        \hline
    \end{tabular}
\end{table}

To quantify the quality of the fits, Table~\ref{tab:chi2} presents the $\chi^2$ values divided by the number of data points, $\chi^2 / N_{\mathrm{pts}}$, obtained for each experimental dataset included in the analysis. The last row shows the total $ \chi^2 $ divided by the number of degrees of freedom, $\chi^2 /\mathrm{d.o.f.}$, for each fit. Note that in this table the dataset of Volmer 2001~\cite{JeffersonLabFpi:2000nlc} also includes the reanalyzed data point of Brauel 1979~\cite{Brauel:1979zk}. Overall, all three analyses provide an acceptable description of the data, with no significant differences observed. This indicates that our ansatz for the GPDs in Eq.~\eqref{Eq3} is not strongly dependent on the choice of PDFs, which is encouraging. The use of the MAP23 PDFs (Set~3) leads to a slightly smaller $\chi^2 /\mathrm{d.o.f.}$, and we therefore adopt this set as our final GPD set. For this fit, the total $\chi^2$ divided by the number of data points for the pion electroproduction and elastic pion scattering processes is $47.260 / 40$ and $113.884 / 101$, respectively.
\begin{table}[!htb]
    \centering
    \caption{$\chi^2$ values per number of data points, $\chi^2 / N_{\mathrm{pts}}$, obtained for each dataset in the three analyses. The last row shows the total $ \chi^2 $ divided by the number of degrees of freedom, $\chi^2 /\mathrm{d.o.f.}$, for each fit.}
    \label{tab:chi2}
    \begin{tabular}{l@{\hskip 0.7cm}c@{\hskip 0.7cm}c@{\hskip 0.7cm}c}
        \hline
        \textbf{Dataset} & \textbf{Set~1} & \textbf{Set~2} & \textbf{Set~3} \\
        \hline
        Brown 1973~\cite{Brown:1973wr}   & $ 2.576 / 5 $ & $ 2.734 / 5 $ & $ 2.885 / 5 $ \\
        Ackermann 1978~\cite{Ackermann:1977rp} & $ 0.036 / 1 $ & $ 0.055 / 1 $ & $ 0.068 / 1 $ \\
        Bebek 1978~\cite{Bebek:1977pe}   & $ 31.372 / 21 $ & $ 31.282 / 21 $ & $ 30.564 / 21 $ \\
        Volmer 2001~\cite{JeffersonLabFpi:2000nlc}  & $ 3.971 / 5 $ & $ 4.058 / 5 $ & $ 4.205 / 5 $ \\
        Huber 2008~\cite{JeffersonLab:2008jve}   & $ 10.145 / 8 $ & $ 9.915 / 8 $ & $ 9.538 / 8 $ \\
        Adylov 1977~\cite{Adylov:1977kj}  & $ 9.449 / 22 $ & $ 9.455 / 22 $ & $ 9.488 / 22 $ \\
        Dally 1981~\cite{Dally:1981ur}   & $ 43.419 / 20 $ & $ 43.202 / 20 $ & $ 43.032 / 20 $ \\
        Dally 1982~\cite{Dally:1982zk}   & $ 9.139 / 14 $ & $ 9.159 / 14 $ & $ 9.140 / 14 $ \\
        Amendolia 1986~\cite{NA7:1986vav} & $ 51.474 / 45 $ & $ 52.656 / 45 $ & $ 52.224 / 45 $ \\
        \hline
        \textbf{Total} & $ 161.581 / 139 $ & $ 162.516 / 139 $ & $ 161.144 / 139 $ \\
        \hline
    \end{tabular}
\end{table}

\begin{figure}[!htb]
    \centering
    \includegraphics[width=0.9\textwidth]{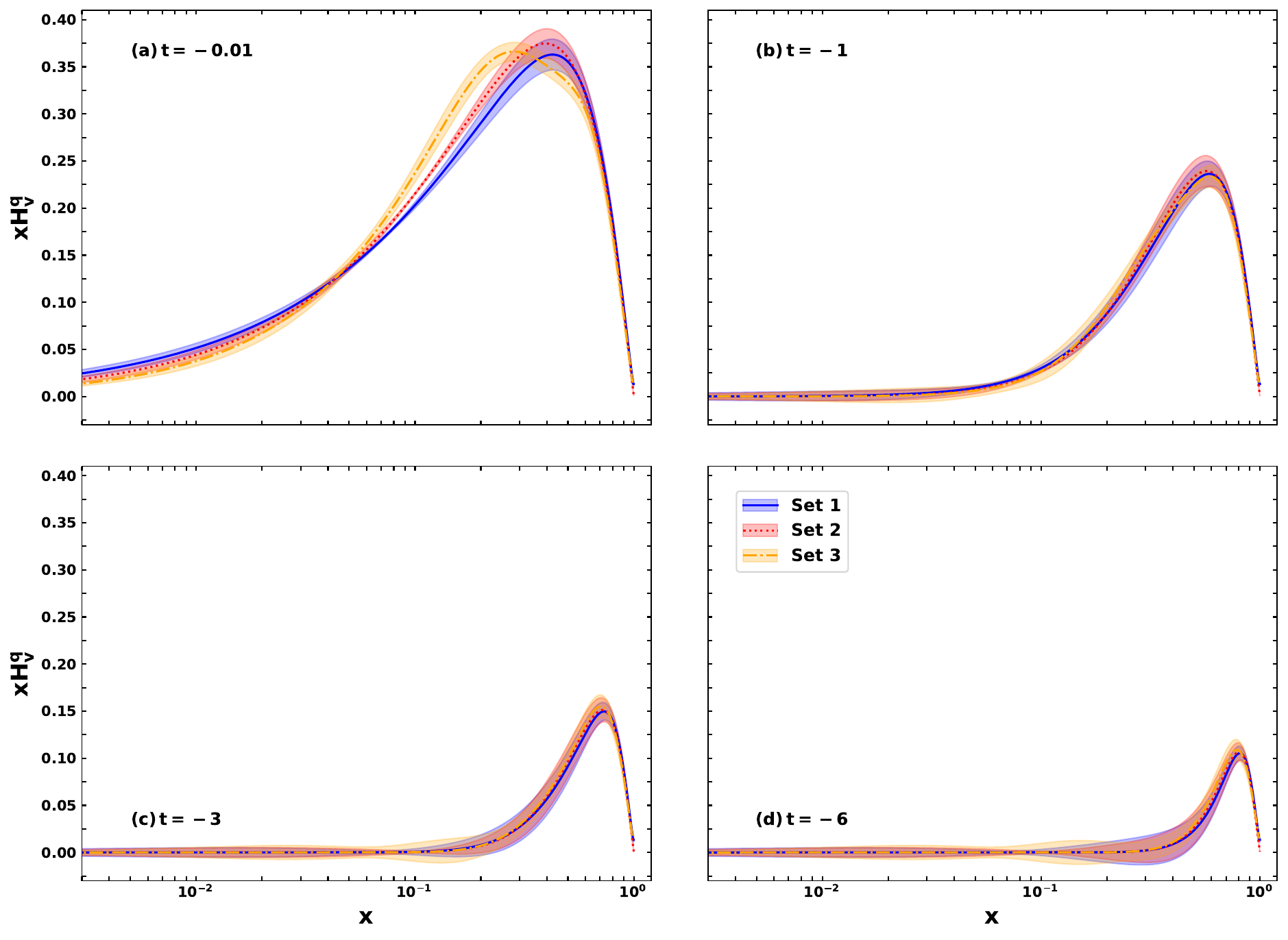}
    \caption{Comparison of the extracted valence pion GPDs, $xH_v^q(x,t,Q^2)$, corresponding to Set~1, Set~2, and Set~3, obtained using the xFitter, JAM21, and MAP23 PDFs, respectively, shown at $Q^2 = 4~\text{GeV}^2$ for representative values of $t$.}
    \label{fig:GPDs}
\end{figure}

In Fig.~\ref{fig:GPDs}, we compare the extracted valence pion GPDs, 
$xH_v^q(x,t,Q^2,\xi=0)$, from the three analyses at $Q^2 = 4~\text{GeV}^2$ 
for representative values of $t$, namely $t = -0.01, -1, -3,$ and $-6~\text{GeV}^2$. Note the uncertainties also include the corresponding ones from the pion PDFs. As expected, the GPDs are suppressed as the absolute value of $t$ increases. The general $x$-dependence is similar across all fits, with some differences emerging primarily in the small-$|t|$ region, where the underlying pion PDFs differ the most. These results provide further evidence that our GPD ansatz, and thus the extracted universal pion GPD, is only weakly sensitive to the choice of pion PDFs.

\begin{figure}[!htb]
    \centering
    \includegraphics[width=0.7\textwidth]{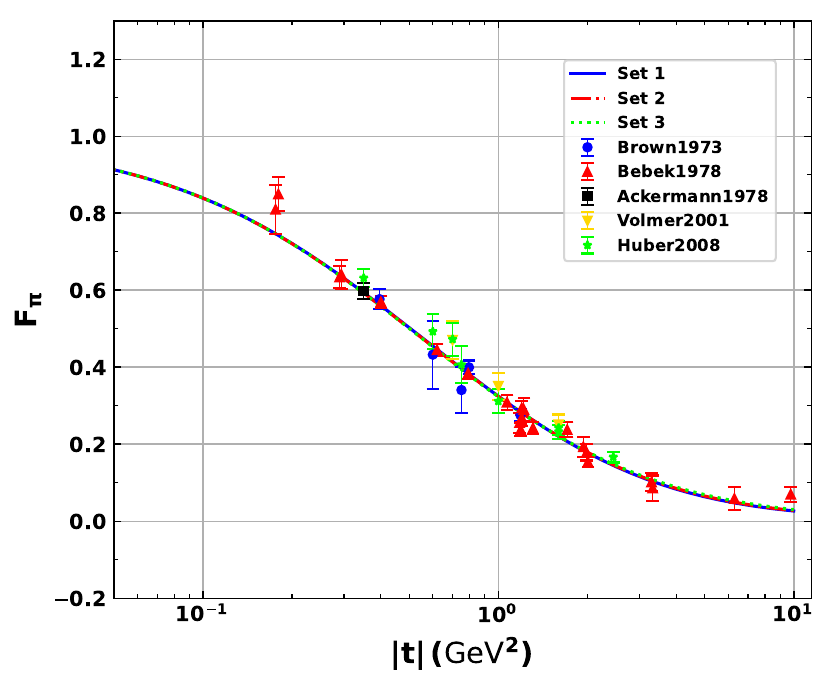}
    \caption{Comparison of the experimental data for the pion electromagnetic form factor, $F_\pi(t)$, from pion electroproduction with the theoretical predictions obtained from the three fits.}
    \label{fig:Fpi}
\end{figure}

\begin{figure}[!htb]
    \centering
    \includegraphics[width=0.7\textwidth]{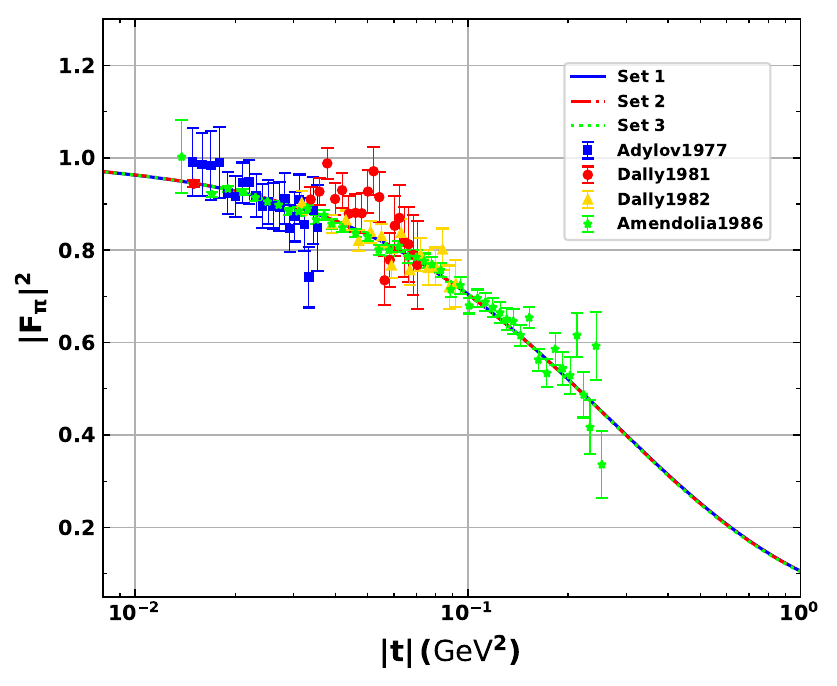}
    \caption{Comparison of the experimental data for the squared pion electromagnetic form factor, $|F_\pi(t)|^2$, from elastic pion scattering with the theoretical predictions obtained from the three fits.}
    \label{fig:F2pi}
\end{figure}

Figures.~\ref{fig:Fpi} and~\ref{fig:F2pi} show the comparison between the theoretical predictions for the pion form factor, $F_\pi(t)$, and its absolute square, $|F_\pi(t)|^2$, obtained from the extracted GPDs, with the experimental measurements from pion electroproduction and elastic scattering. All three fits provide a satisfactory description of the data across the full $t$ range. As can be seen, all GPD sets lead to essentially the same predictions. As mentioned earlier, since the analysis using MAP23 PDFs results in a slightly smaller $\chi^2$, we take it as our final result.

\begin{figure}[!htb]
    \centering
    \includegraphics[width=0.7\textwidth]{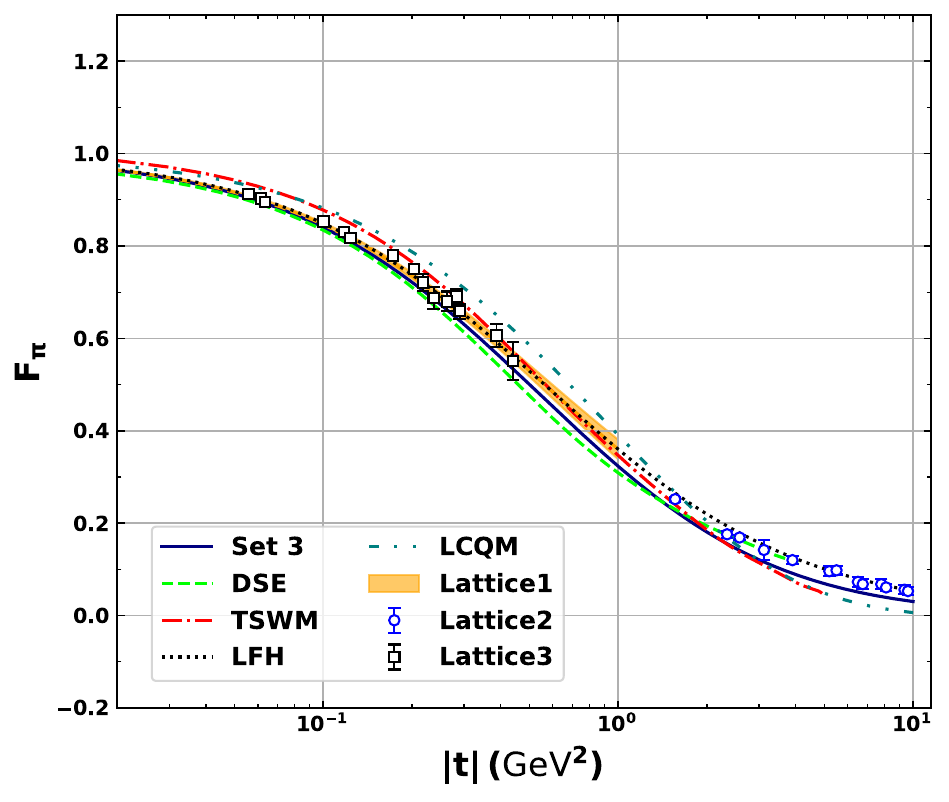}
    \caption{Comparison of the pion electromagnetic form factor, $F_\pi(t)$, obtained from the present analysis (Set~3), with representative theoretical predictions from the DSE approach~\cite{Chang:2013nia}, LFH~\cite{Brodsky:2007hb}, LCQM~\cite{Puhan:2025pfs}, and TSWM of holographic QCD~\cite{Nasibova:2025wnw}, together with recent lattice QCD results~\cite{Wang:2020nbf,Ding:2024lfj,Gao:2021xsm}.}
    \label{fig:FpiComp}
\end{figure}

Figure~\ref{fig:FpiComp} shows a comparison between the theoretical predictions for the pion electromagnetic form factor, $F_\pi(t)$, obtained from the present analysis (Set~3), and representative results from several established theoretical approaches, including the DSE framework~\cite{Chang:2013nia}, LFH~\cite{Brodsky:2007hb}, LCQM~\cite{Puhan:2025pfs}, and the thermal soft-wall model (TSWM) of holographic QCD~\cite{Nasibova:2025wnw}, together with recent lattice QCD results labeled as Lattice~1~\cite{Wang:2020nbf}, Lattice~2~\cite{Ding:2024lfj}, and Lattice~3~\cite{Gao:2021xsm}. 
Overall, our prediction agrees very well with these approaches in whole $|t|$ region, particularly with the DSE, LFH, and lattice QCD calculations at small $|t|$ region. At larger $|t|$, our result lies within the spread of existing model predictions and follows the average trend of the different calculations. This comparison demonstrates that the GPD-based extraction obtained in the present work is fully consistent with a broad range of modern theoretical frameworks.

In addition to the pion electromagnetic FF itself, it is instructive to compare the results with the scaled quantity $|t| F_\pi(t)$. This observable highlights the asymptotic behavior of the FF and provides a direct connection to perturbative QCD (pQCD) predictions~\cite{Bakulev:2004cu}. Experimentally, measurements of $|t| F_\pi(t)$ from Jefferson Lab (JLab)~\cite{JeffersonLabFpi:2000nlc,JeffersonLab:2008jve} have provided valuable insights into the transition region between the nonperturbative and perturbative regimes of QCD. In addition, results from lattice QCD and other theoretical and phenomenological approaches are available for this quantity, allowing for a direct comparison between theory, phenomenology, and experiment.

\begin{figure}[!htb]
    \centering
    \includegraphics[width=0.7\textwidth]{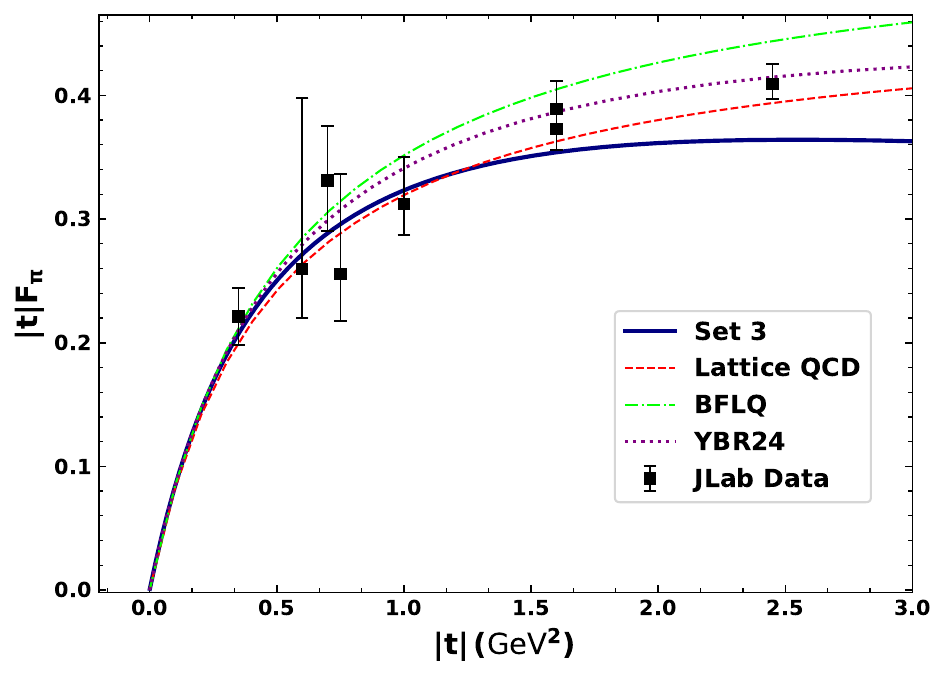}
    \caption{Comparison of the experimental data for $|t| F_\pi(t)$ from JLab~\cite{JeffersonLab:2008jve} with theoretical predictions from our analysis (Set~3), as well as results from lattice QCD~\cite{QCDSFUKQCD:2006gmg}, BLFQ~\cite{Adhikari:2021jrh}, and YBR24~\cite{Yao:2024drm}.}
    \label{fig:tFpi}
\end{figure}

Figure~\ref{fig:tFpi} presents our theoretical predictions for $|t| F_\pi(t)$ obtained from the final GPD set (Set~3), together with the experimental measurements from JLab~\cite{JeffersonLab:2008jve} as well as results from lattice QCD~\cite{QCDSFUKQCD:2006gmg}, LFH by the BLFQ Collaboration~\cite{Adhikari:2021jrh}, and parameter-free prediction of Ref.~\cite{Yao:2024drm} obtained using a symmetry-preserving truncation of the quantum field equations describing hadron properties (labeled as YBR24). The agreement between our predictions and the data is very good, except in the large-$|t|$ region. Note that the JLab $F_\pi(t)$ data were already included in our analyses, labeled as Huber 2008 in Tables~\ref{tab:datasets} and~\ref{tab:chi2}, with a corresponding $\chi^2$ of 9.538 for 8 data points (for Set~3 in Table~\ref{tab:chi2}), indicating excellent fit quality. The minor deviation observed in Fig.~\ref{fig:tFpi} at large $|t|$ reflects the sensitivity of the scaled quantity $|t| F_\pi(t)$, and hence, highlights its role as a precision observable. From another point of view, it is also possible that the JLab data may require refinement with future precise measurements, especially since pQCD predicts that $|t| F_\pi(t)$ should approach a constant value at large momentum transfers. Such a behavior well reproduced by our results.

%

\section{Summary and conclusion}\label{sec:six} 
 
In this work, we have performed a global QCD analysis of the pion electromagnetic FF data to extract the valence pion GPDs at zero skewness. The analysis was carried out using three different sets of pion PDFs, namely xFitter~\cite{Novikov:2020snp}, JAM21~\cite{Barry:2021osv}, and MAP23~\cite{Pasquini:2023aaf}, which were used to construct the GPD ansatz. We carefully examined the dependence of the extracted GPDs on the choice of the underlying pion PDF and performed a parametrization scan procedure to determine the optimum values of the model parameters.

The profile function parameters of the GPDs were determined through a $\chi^2$ minimization. The total number of experimental data points included in the analyses is 141, comprising 40 points from pion electroproduction and 101 points from elastic pion scattering. By following standard assumptions on the pion PDFs and their valence quark symmetry, and employing a parametrization scan procedure, we reduced the number of free parameters and found that only the parameters $\alpha'$ and $A$ were sufficient to achieve a satisfactory description of the data. Our results indicate that the extracted parameters exhibit only a mild dependence on the choice of PDF set, while the overall trends remain consistent across the three analyses.

A detailed comparison of the extracted valence pion GPDs from analyses with different input PDFs shows that the $x$-dependence is very similar in all fits. Only minor differences are seen at small $t$ which reflects variations in the input PDFs. The theoretical predictions for the pion electromagnetic form factor, $F_\pi(t)$, and its absolute square, $|F_\pi(t)|^2$, are in good agreement with the experimental measurements. This good agreement demonstrates the reliability of our GPD ansatz. Among the three analyses, the one using the MAP23 PDFs (Set~3) leads to a slightly smaller $\chi^2/\mathrm{d.o.f.}$ and is therefore adopted as our final GPD set.

In summary, our study provides a comprehensive determination of the valence pion GPDs at $Q^2 = 4~\text{GeV}^2$ and $ \xi=0 $, showing that the choice of pion PDFs has only a minor effect on the extracted distributions. These results represent a significant step forward in understanding the internal structure of the pion within QCD and provide a solid foundation for future phenomenological studies, such as investigations of the pion charge radius, pion tomography, and pion mechanical properties, similar to recent analyses performed for the proton~\cite{Goharipour:2024mbk,Hashamipour:2020kip,Adhikari:2021jrh,Goharipour:2025lep}.

%
\section*{ACKNOWLEDGEMENTS}
F. Irani and K. Azizi are tankful to Iran National Science Foundation (INSF) for financial support provided for this research under Grant No. 4033039.

%

%

\bibliographystyle{apsrev4-1}
\bibliography{article}

@article{Yndurain:2002ud,
    author = "Yndurain, F. J.",
    title = "{Low-energy pion physics}",
    eprint = "hep-ph/0212282",
    archivePrefix = "arXiv",
    reportNumber = "FTUAM-02-28",
    month = "12",
    year = "2002"
}

@article{Horn:2016rip,
    author = "Horn, Tanja and Roberts, Craig D.",
    title = "{The pion: an enigma within the Standard Model}",
    eprint = "1602.04016",
    archivePrefix = "arXiv",
    primaryClass = "nucl-th",
    doi = "10.1088/0954-3899/43/7/073001",
    journal = "J. Phys. G",
    volume = "43",
    number = "7",
    pages = "073001",
    year = "2016"
}

@article{Aguilar:2019teb,
    author = "Aguilar, Arlene C. and others",
    title = "{Pion and Kaon Structure at the Electron-Ion Collider}",
    eprint = "1907.08218",
    archivePrefix = "arXiv",
    primaryClass = "nucl-ex",
    reportNumber = "NJU-INP 001/19",
    doi = "10.1140/epja/i2019-12885-0",
    journal = "Eur. Phys. J. A",
    volume = "55",
    number = "10",
    pages = "190",
    year = "2019"
}

@article{Ananthanarayan:2022wsl,
    author = "Ananthanarayan, B.",
    title = "{Pions: the original Nambu{\textendash}Goldstone bosons: An introduction and precision pion physics}",
    doi = "10.1140/epjs/s11734-022-00443-7",
    journal = "Eur. Phys. J. ST",
    volume = "231",
    number = "2",
    pages = "91--102",
    year = "2022"
}

@article{Holt:2010vj,
    author = "Holt, Roy J. and Roberts, Craig D.",
    title = "{Distribution Functions of the Nucleon and Pion in the Valence Region}",
    eprint = "1002.4666",
    archivePrefix = "arXiv",
    primaryClass = "nucl-th",
    doi = "10.1103/RevModPhys.82.2991",
    journal = "Rev. Mod. Phys.",
    volume = "82",
    pages = "2991--3044",
    year = "2010"
}

@article{Ethier:2020way,
    author = "Ethier, Jacob J. and Nocera, Emanuele R.",
    title = "{Parton Distributions in Nucleons and Nuclei}",
    eprint = "2001.07722",
    archivePrefix = "arXiv",
    primaryClass = "hep-ph",
    reportNumber = "Nikhef/2020-003",
    doi = "10.1146/annurev-nucl-011720-042725",
    journal = "Ann. Rev. Nucl. Part. Sci.",
    volume = "70",
    pages = "43--76",
    year = "2020"
}

@inproceedings{Lorce:2025aqp,
    author = "Lorc{\'e}, C{\'e}dric and Metz, Andreas and Pasquini, Barbara and Schweitzer, Peter",
    title = "{Parton Distribution Functions and their Generalizations}",
    eprint = "2507.12664",
    archivePrefix = "arXiv",
    primaryClass = "hep-ph",
    month = "7",
    year = "2025"
}

@article{H1:2015ubc,
    author = "Abramowicz, H. and others",
    collaboration = "H1, ZEUS",
    title = "{Combination of measurements of inclusive deep inelastic ${e^{\pm }p}$ scattering cross sections and QCD analysis of HERA data}",
    eprint = "1506.06042",
    archivePrefix = "arXiv",
    primaryClass = "hep-ex",
    reportNumber = "DESY-15-039",
    doi = "10.1140/epjc/s10052-015-3710-4",
    journal = "Eur. Phys. J. C",
    volume = "75",
    number = "12",
    pages = "580",
    year = "2015"
}

@article{Accardi:2016qay,
    author = "Accardi, A. and Brady, L. T. and Melnitchouk, W. and Owens, J. F. and Sato, N.",
    title = "{Constraints on large-$x$ parton distributions from new weak boson production and deep-inelastic scattering data}",
    eprint = "1602.03154",
    archivePrefix = "arXiv",
    primaryClass = "hep-ph",
    reportNumber = "JLAB-THY-16-2215",
    doi = "10.1103/PhysRevD.93.114017",
    journal = "Phys. Rev. D",
    volume = "93",
    number = "11",
    pages = "114017",
    year = "2016"
}

@article{Alekhin:2017kpj,
    author = {Alekhin, S. and Bl{\"u}mlein, J. and Moch, S. and Placakyte, R.},
    title = "{Parton distribution functions, $\alpha_s$, and heavy-quark masses for LHC Run II}",
    eprint = "1701.05838",
    archivePrefix = "arXiv",
    primaryClass = "hep-ph",
    reportNumber = "DESY-16-179, DO-TH-16-13",
    doi = "10.1103/PhysRevD.96.014011",
    journal = "Phys. Rev. D",
    volume = "96",
    number = "1",
    pages = "014011",
    year = "2017"
}

@article{Hou:2019efy,
    author = "Hou, Tie-Jiun and others",
    title = "{New CTEQ global analysis of quantum chromodynamics with high-precision data from the LHC}",
    eprint = "1912.10053",
    archivePrefix = "arXiv",
    primaryClass = "hep-ph",
    reportNumber = "MSUHEP-19-025, PITT-PACC-1911, SMU-HEP-19-03",
    doi = "10.1103/PhysRevD.103.014013",
    journal = "Phys. Rev. D",
    volume = "103",
    number = "1",
    pages = "014013",
    year = "2021"
}

@article{Bailey:2020ooq,
    author = "Bailey, S. and Cridge, T. and Harland-Lang, L. A. and Martin, A. D. and Thorne, R. S.",
    title = "{Parton distributions from LHC, HERA, Tevatron and fixed target data: MSHT20 PDFs}",
    eprint = "2012.04684",
    archivePrefix = "arXiv",
    primaryClass = "hep-ph",
    reportNumber = "IPPP/20/58",
    doi = "10.1140/epjc/s10052-021-09057-0",
    journal = "Eur. Phys. J. C",
    volume = "81",
    number = "4",
    pages = "341",
    year = "2021"
}

@article{ATLAS:2021qnl,
    author = "Aad, Georges and others",
    collaboration = "ATLAS",
    title = "{Determination of the parton distribution functions of the proton from ATLAS measurements of differential W$^{±}$ and Z boson production in association with jets}",
    eprint = "2101.05095",
    archivePrefix = "arXiv",
    primaryClass = "hep-ex",
    reportNumber = "CERN-EP-2020-237",
    doi = "10.1007/JHEP07(2021)223",
    journal = "JHEP",
    volume = "07",
    pages = "223",
    year = "2021"
}

@article{NNPDF:2021njg,
    author = "Ball, Richard D. and others",
    collaboration = "NNPDF",
    title = "{The path to proton structure at 1{\%} accuracy}",
    eprint = "2109.02653",
    archivePrefix = "arXiv",
    primaryClass = "hep-ph",
    reportNumber = "Edinburgh 2021/12, Nikhef-2021-013, TIF-UNIMI-2021-11",
    doi = "10.1140/epjc/s10052-022-10328-7",
    journal = "Eur. Phys. J. C",
    volume = "82",
    number = "5",
    pages = "428",
    year = "2022"
}

@article{PDF4LHCWorkingGroup:2022cjn,
    author = "Ball, Richard D. and others",
    collaboration = "PDF4LHC Working Group",
    title = "{The PDF4LHC21 combination of global PDF fits for the LHC Run III}",
    eprint = "2203.05506",
    archivePrefix = "arXiv",
    primaryClass = "hep-ph",
    reportNumber = "Edinburgh 2021/31, FERMILAB-PUB-22-121-QIS-SCD-T, MSUHEP-22-010,
  Nikhef 2021-033, SMU-HEP-22-01",
    doi = "10.1088/1361-6471/ac7216",
    journal = "J. Phys. G",
    volume = "49",
    number = "8",
    pages = "080501",
    year = "2022"
}

@article{deFlorian:2008mr,
    author = "de Florian, Daniel and Sassot, Rodolfo and Stratmann, Marco and Vogelsang, Werner",
    title = "{Global Analysis of Helicity Parton Densities and Their Uncertainties}",
    eprint = "0804.0422",
    archivePrefix = "arXiv",
    primaryClass = "hep-ph",
    reportNumber = "BNL-NT-08-8",
    doi = "10.1103/PhysRevLett.101.072001",
    journal = "Phys. Rev. Lett.",
    volume = "101",
    pages = "072001",
    year = "2008"
}

@article{Leader:2010rb,
    author = "Leader, Elliot and Sidorov, Aleksander V. and Stamenov, Dimiter B.",
    title = "{Determination of Polarized PDFs from a QCD Analysis of Inclusive and Semi-inclusive Deep Inelastic Scattering Data}",
    eprint = "1010.0574",
    archivePrefix = "arXiv",
    primaryClass = "hep-ph",
    doi = "10.1103/PhysRevD.82.114018",
    journal = "Phys. Rev. D",
    volume = "82",
    pages = "114018",
    year = "2010"
}

@article{Blumlein:2010rn,
    author = "Blumlein, Johannes and Bottcher, Helmut",
    title = "{QCD Analysis of Polarized Deep Inelastic Scattering Data}",
    eprint = "1005.3113",
    archivePrefix = "arXiv",
    primaryClass = "hep-ph",
    reportNumber = "DESY-09-131, SFB-CPP-10-032",
    doi = "10.1016/j.nuclphysb.2010.08.005",
    journal = "Nucl. Phys. B",
    volume = "841",
    pages = "205--230",
    year = "2010"
}

@article{Nocera:2014gqa,
    author = "Nocera, Emanuele R. and Ball, Richard D. and Forte, Stefano and Ridolfi, Giovanni and Rojo, Juan",
    collaboration = "NNPDF",
    title = "{A first unbiased global determination of polarized PDFs and their uncertainties}",
    eprint = "1406.5539",
    archivePrefix = "arXiv",
    primaryClass = "hep-ph",
    reportNumber = "CERN-PH-TH-2014-106, IFUN-1028-FT, EDINBURGH-14-11, OUTP-14-06P",
    doi = "10.1016/j.nuclphysb.2014.08.008",
    journal = "Nucl. Phys. B",
    volume = "887",
    pages = "276--308",
    year = "2014"
}

@article{Jimenez-Delgado:2014xza,
    author = "Jimenez-Delgado, P. and Avakian, H. and Melnitchouk, W.",
    collaboration = "Jefferson Lab Angular Momentum (JAM)",
    title = "{Constraints on spin-dependent parton distributions at large x from global QCD analysis}",
    eprint = "1403.3355",
    archivePrefix = "arXiv",
    primaryClass = "hep-ph",
    reportNumber = "JLAB-THY-14-1856",
    doi = "10.1016/j.physletb.2014.09.049",
    journal = "Phys. Lett. B",
    volume = "738",
    pages = "263--267",
    year = "2014"
}

@article{Salajegheh:2018hfs,
    author = "Salajegheh, Maral and Moosavi Nejad, S. Mohammad and Nejad, Moosavi and Khanpour, Hamzeh and Atashbar Tehrani, S.",
    title = "{Analytical approaches to the determination of spin-dependent parton distribution functions at NNLO approximation}",
    eprint = "1801.04471",
    archivePrefix = "arXiv",
    primaryClass = "hep-ph",
    doi = "10.1103/PhysRevC.97.055201",
    journal = "Phys. Rev. C",
    volume = "97",
    number = "5",
    pages = "055201",
    year = "2018"
}

@article{Adamiak:2023yhz,
    author = "Adamiak, Daniel and Baldonado, Nicholas and Kovchegov, Yuri V. and Melnitchouk, W. and Pitonyak, Daniel and Sato, Nobuo and Sievert, Matthew D. and Tarasov, Andrey and Tawabutr, Yossathorn",
    collaboration = "Jefferson Lab Angular Momentum (JAM)",
    title = "{Global analysis of polarized DIS and SIDIS data with improved small-x helicity evolution}",
    eprint = "2308.07461",
    archivePrefix = "arXiv",
    primaryClass = "hep-ph",
    reportNumber = "JLAB-THY-23-3896",
    doi = "10.1103/PhysRevD.108.114007",
    journal = "Phys. Rev. D",
    volume = "108",
    number = "11",
    pages = "114007",
    year = "2023"
}

@article{Bertone:2024taw,
    author = "Bertone, Valerio and Chiefa, Amedeo and Nocera, Emanuele R.",
    collaboration = "MAP (Multi-dimensional Analyses of Partonic distributions)",
    title = "{Helicity-dependent parton distribution functions at next-to-next-to-leading order accuracy from inclusive and semi-inclusive deep-inelastic scattering data}",
    eprint = "2404.04712",
    archivePrefix = "arXiv",
    primaryClass = "hep-ph",
    doi = "10.1016/j.physletb.2025.139497",
    journal = "Phys. Lett. B",
    volume = "865",
    pages = "139497",
    year = "2025"
}

@article{Borsa:2024mss,
    author = "Borsa, Ignacio and Stratmann, Marco and Vogelsang, Werner and de Florian, Daniel and Sassot, Rodolfo",
    title = "{Next-to-Next-to-Leading Order Global Analysis of Polarized Parton Distribution Functions}",
    eprint = "2407.11635",
    archivePrefix = "arXiv",
    primaryClass = "hep-ph",
    doi = "10.1103/PhysRevLett.133.151901",
    journal = "Phys. Rev. Lett.",
    volume = "133",
    number = "15",
    pages = "151901",
    year = "2024"
}

@article{Cruz-Martinez:2025ahf,
    author = "Cruz-Martinez, Juan and Hasenack, Toon and Hekhorn, Felix and Magni, Giacomo and Nocera, Emanuele R. and Rabemananjara, Tanjona R. and Rojo, Juan and Sharma, Tanishq and van Seeventer, Gijs",
    title = "{NNPDFpol2.0: a global determination of polarised PDFs and their uncertainties at next-to-next-to-leading order}",
    eprint = "2503.11814",
    archivePrefix = "arXiv",
    primaryClass = "hep-ph",
    doi = "10.1007/JHEP07(2025)168",
    journal = "JHEP",
    volume = "07",
    pages = "168",
    year = "2025"
}

@article{Gluck:1991ey,
    author = "Gluck, M. and Reya, E. and Vogt, A.",
    title = "{Pionic parton distributions}",
    reportNumber = "DO-TH-91-16",
    doi = "10.1007/BF01559743",
    journal = "Z. Phys. C",
    volume = "53",
    pages = "651--656",
    year = "1992"
}

@article{Wijesooriya:2005ir,
    author = "Wijesooriya, K. and Reimer, P. E. and Holt, R. J.",
    title = "{The pion parton distribution function in the valence region}",
    eprint = "nucl-ex/0509012",
    archivePrefix = "arXiv",
    doi = "10.1103/PhysRevC.72.065203",
    journal = "Phys. Rev. C",
    volume = "72",
    pages = "065203",
    year = "2005"
}

@article{Aicher:2010cb,
    author = "Aicher, Matthias and Schafer, Andreas and Vogelsang, Werner",
    title = "{Soft-gluon resummation and the valence parton distribution function of the pion}",
    eprint = "1009.2481",
    archivePrefix = "arXiv",
    primaryClass = "hep-ph",
    doi = "10.1103/PhysRevLett.105.252003",
    journal = "Phys. Rev. Lett.",
    volume = "105",
    pages = "252003",
    year = "2010"
}

@article{Han:2020vjp,
    author = "Han, Chengdong and Xie, Gang and Wang, Rong and Chen, Xurong",
    title = "{An Analysis of Parton Distribution Functions of the Pion and the Kaon with the Maximum Entropy Input}",
    eprint = "2010.14284",
    archivePrefix = "arXiv",
    primaryClass = "hep-ph",
    doi = "10.1140/epjc/s10052-021-09087-8",
    journal = "Eur. Phys. J. C",
    volume = "81",
    number = "4",
    pages = "302",
    year = "2021"
}

@article{Novikov:2020snp,
    author = "Novikov, Ivan and others",
    title = "{Parton Distribution Functions of the Charged Pion Within The xFitter Framework}",
    eprint = "2002.02902",
    archivePrefix = "arXiv",
    primaryClass = "hep-ph",
    reportNumber = "DESY-20-013, DESY 20-013",
    doi = "10.1103/PhysRevD.102.014040",
    journal = "Phys. Rev. D",
    volume = "102",
    number = "1",
    pages = "014040",
    year = "2020"
}

@article{Barry:2018ort,
    author = "Barry, P. C. and Sato, N. and Melnitchouk, W. and Ji, Chueng-Ryong",
    title = "{First Monte Carlo Global QCD Analysis of Pion Parton Distributions}",
    eprint = "1804.01965",
    archivePrefix = "arXiv",
    primaryClass = "hep-ph",
    reportNumber = "JLAB-THY-18-2678",
    doi = "10.1103/PhysRevLett.121.152001",
    journal = "Phys. Rev. Lett.",
    volume = "121",
    number = "15",
    pages = "152001",
    year = "2018"
}

@article{Barry:2021osv,
    author = "Barry, P. C. and Ji, Chueng-Ryong and Sato, N. and Melnitchouk, W.",
    collaboration = "Jefferson Lab Angular Momentum (JAM)",
    title = "{Global QCD Analysis of Pion Parton Distributions with Threshold Resummation}",
    eprint = "2108.05822",
    archivePrefix = "arXiv",
    primaryClass = "hep-ph",
    reportNumber = "JLAB-THY-21-3482",
    doi = "10.1103/PhysRevLett.127.232001",
    journal = "Phys. Rev. Lett.",
    volume = "127",
    number = "23",
    pages = "232001",
    year = "2021"
}

@article{Barry:2025wjx,
    author = "Barry, P. C. and Ji, Chueng-Ryong and Melnitchouk, W. and Sato, N. and Steffens, Fernanda",
    collaboration = "JAM",
    title = "{First simultaneous global QCD analysis of kaon and pion parton distributions with lattice QCD constraints}",
    eprint = "2510.11979",
    archivePrefix = "arXiv",
    primaryClass = "hep-ph",
    reportNumber = "JLAB-THY-25-4569",
    month = "10",
    year = "2025"
}

@article{Pasquini:2023aaf,
    author = "Pasquini, Barbara and Rodini, Simone and Venturini, Simone",
    collaboration = "MAP (Multi-dimensional Analyses of Partonic distributions)",
    title = "{Valence quark, sea, and gluon content of the pion from the parton distribution functions and the electromagnetic form factor}",
    eprint = "2303.01789",
    archivePrefix = "arXiv",
    primaryClass = "hep-ph",
    doi = "10.1103/PhysRevD.107.114023",
    journal = "Phys. Rev. D",
    volume = "107",
    number = "11",
    pages = "114023",
    year = "2023"
}

@article{Kotz:2023pbu,
    author = "Kotz, Lucas and Courtoy, Aurore and Nadolsky, Pavel and Olness, Fredrick and Ponce-Chavez, Maximiliano",
    title = "{Analysis of parton distributions in a pion with B{\'e}zier parametrizations}",
    eprint = "2311.08447",
    archivePrefix = "arXiv",
    primaryClass = "hep-ph",
    reportNumber = "FERMILAB-PUB-23-695-V, SMU-HEP-23-04",
    doi = "10.1103/PhysRevD.109.074027",
    journal = "Phys. Rev. D",
    volume = "109",
    number = "7",
    pages = "074027",
    year = "2024"
}

@article{Kotz:2025lio,
    author = "Kotz, Lucas and Courtoy, Aurore and Nadolsky, Pavel and Ponce-Chavez, Maximiliano",
    title = "{Epistemic and nuclear uncertainties for the parton distributions of the pion}",
    eprint = "2505.13594",
    archivePrefix = "arXiv",
    primaryClass = "hep-ph",
    doi = "10.1103/h2vn-1wxp",
    journal = "Phys. Rev. D",
    volume = "112",
    number = "7",
    pages = "L071502",
    year = "2025"
}

@article{Good:2025nny,
    author = "Good, William and Barry, Patrick C. and Lin, Huey-Wen and Melnitchouk, W. and NieMiera, Alex and Sato, Nobuo",
    title = "{Pionic gluons from global QCD analysis of experimental and lattice data}",
    eprint = "2507.22730",
    archivePrefix = "arXiv",
    primaryClass = "hep-ph",
    reportNumber = "MSUHEP-25-014, JLAB-THY-25-4417",
    month = "7",
    year = "2025"
}

@article{Gluck:1999xe,
    author = "Gluck, M. and Reya, E. and Schienbein, I.",
    title = "{Pionic parton distributions revisited}",
    eprint = "hep-ph/9903288",
    archivePrefix = "arXiv",
    reportNumber = "DO-TH-99-01",
    doi = "10.1007/s100529900124",
    journal = "Eur. Phys. J. C",
    volume = "10",
    pages = "313--317",
    year = "1999"
}

@article{Abdel-Rehim:2015owa,
    author = "Abdel-Rehim, A. and others",
    title = "{Nucleon and pion structure with lattice QCD simulations at physical value of the pion mass}",
    eprint = "1507.04936",
    archivePrefix = "arXiv",
    primaryClass = "hep-lat",
    doi = "10.1103/PhysRevD.92.114513",
    journal = "Phys. Rev. D",
    volume = "92",
    number = "11",
    pages = "114513",
    year = "2015",
    note = "[Erratum: Phys.Rev.D 93, 039904 (2016)]"
}

@article{Cui:2022bxn,
    author = "Cui, Z. -F. and Ding, Minghui and Morgado, J. M. and Raya, K. and Binosi, D. and Chang, L. and De Soto, F. and Roberts, C. D. and Rodr{\'\i}guez-Quintero, J. and Schmidt, S. M.",
    title = "{Emergence of pion parton distributions}",
    eprint = "2201.00884",
    archivePrefix = "arXiv",
    primaryClass = "hep-ph",
    reportNumber = "NJU-INP 054/22",
    doi = "10.1103/PhysRevD.105.L091502",
    journal = "Phys. Rev. D",
    volume = "105",
    number = "9",
    pages = "L091502",
    year = "2022"
}

@article{Hutauruk:2023ccw,
    author = "Hutauruk, Parada T. P. and Nam, Seung-il",
    title = "{Updated analyses of gluon distribution functions for the pion and kaon from the gauge-invariant nonlocal chiral quark model}",
    eprint = "2302.05566",
    archivePrefix = "arXiv",
    primaryClass = "hep-ph",
    doi = "10.1103/PhysRevD.109.054040",
    journal = "Phys. Rev. D",
    volume = "109",
    number = "5",
    pages = "054040",
    year = "2024"
}

@article{Lan:2019rba,
    author = "Lan, Jiangshan and Mondal, Chandan and Jia, Shaoyang and Zhao, Xingbo and Vary, James P.",
    title = "{Pion and kaon parton distribution functions from basis light front quantization and QCD evolution}",
    eprint = "1907.01509",
    archivePrefix = "arXiv",
    primaryClass = "nucl-th",
    doi = "10.1103/PhysRevD.101.034024",
    journal = "Phys. Rev. D",
    volume = "101",
    number = "3",
    pages = "034024",
    year = "2020"
}

@article{Lan:2024ais,
    author = "Lan, Jiangshan and Mondal, Chandan and Zhao, Xingbo and Frederico, Tobias and Vary, James P.",
    title = "{Gluonic contributions to the pion parton distribution functions}",
    eprint = "2406.18878",
    archivePrefix = "arXiv",
    primaryClass = "hep-ph",
    doi = "10.1103/3z39-l1kg",
    journal = "Phys. Rev. D",
    volume = "111",
    number = "11",
    pages = "L111903",
    year = "2025"
}

@article{Chen:2024dhz,
    author = "Chen, Jingxuan and Wang, Xiaopeng and Cai, Yanbing and Chen, Xurong and Wang, Qian",
    title = "{Valence Quark Distributions in Pions: Insights from Tsallis Entropy}",
    eprint = "2408.03068",
    archivePrefix = "arXiv",
    primaryClass = "hep-ph",
    month = "8",
    year = "2024"
}

@article{Maerovitz:2025txk,
    author = "Maerovitz, Joseph and Sosa, Alan and Leon, Christopher and Sargsian, Misak",
    title = "{Pion Valence Structure at Intermediate x in the Residual Field Model}",
    eprint = "2506.15995",
    archivePrefix = "arXiv",
    primaryClass = "hep-ph",
    month = "6",
    year = "2025"
}

@article{Kaur:2025gyr,
    author = "Kaur, Satvir and Mondal, Chandan",
    title = "{Gluon Distributions in the Pion}",
    eprint = "2507.01506",
    archivePrefix = "arXiv",
    primaryClass = "hep-ph",
    month = "7",
    year = "2025"
}

@article{Bopsin:2025vhz,
    author = "Bopsin, Gustavo B. and El-Bennich, Bruno and Krein, Gast{\~a}o and Serna, Fernando E. and da Silveira, Roberto C.",
    title = "{Parton distribution and fragmentation functions with massive gluons}",
    eprint = "2507.12544",
    archivePrefix = "arXiv",
    primaryClass = "hep-ph",
    month = "7",
    year = "2025"
}

@article{Francis:2025rya,
    author = "Francis, Anthony and others",
    title = "{Gradient flow for parton distribution functions: first application to the pion}",
    eprint = "2509.02472",
    archivePrefix = "arXiv",
    primaryClass = "hep-lat",
    month = "9",
    year = "2025"
}

@article{Wang:2025usl,
    author = "Wang, Xiaobin and Chang, Lei and Ding, Minghui and Raya, Khepani and Roberts, Craig D.",
    title = "{Symmetry Constraints on Pion Valence Structure}",
    eprint = "2510.23950",
    archivePrefix = "arXiv",
    primaryClass = "hep-ph",
    reportNumber = "NJU-INP 106-25",
    month = "10",
    year = "2025"
}

@article{Ji:1998pc,
    author = "Ji, Xiang-Dong",
    title = "{Off forward parton distributions}",
    eprint = "hep-ph/9807358",
    archivePrefix = "arXiv",
    reportNumber = "UMD-PP-98-092, DOE-ER-40762-144",
    doi = "10.1088/0954-3899/24/7/002",
    journal = "J. Phys. G",
    volume = "24",
    pages = "1181--1205",
    year = "1998"
}

@article{Goeke:2001tz,
    author = "Goeke, K. and Polyakov, Maxim V. and Vanderhaeghen, M.",
    title = "{Hard exclusive reactions and the structure of hadrons}",
    eprint = "hep-ph/0106012",
    archivePrefix = "arXiv",
    doi = "10.1016/S0146-6410(01)00158-2",
    journal = "Prog. Part. Nucl. Phys.",
    volume = "47",
    pages = "401--515",
    year = "2001"
}

@article{Burkardt:2000za,
    author = "Burkardt, Matthias",
    title = "{Impact parameter dependent parton distributions and off forward parton distributions for zeta ---{\ensuremath{>}} 0}",
    eprint = "hep-ph/0005108",
    archivePrefix = "arXiv",
    doi = "10.1103/PhysRevD.62.071503",
    journal = "Phys. Rev. D",
    volume = "62",
    pages = "071503",
    year = "2000",
    note = "[Erratum: Phys.Rev.D 66, 119903 (2002)]"
}

@article{Burkardt:2002hr,
    author = "Burkardt, Matthias",
    title = "{Impact parameter space interpretation for generalized parton distributions}",
    eprint = "hep-ph/0207047",
    archivePrefix = "arXiv",
    doi = "10.1142/S0217751X03012370",
    journal = "Int. J. Mod. Phys. A",
    volume = "18",
    pages = "173--208",
    year = "2003"
}

@article{Polyakov:2002yz,
    author = "Polyakov, M. V.",
    title = "{Generalized parton distributions and strong forces inside nucleons and nuclei}",
    eprint = "hep-ph/0210165",
    archivePrefix = "arXiv",
    reportNumber = "RUB-TP2-14-02",
    doi = "10.1016/S0370-2693(03)00036-4",
    journal = "Phys. Lett. B",
    volume = "555",
    pages = "57--62",
    year = "2003"
}

@article{Diehl:2003ny,
    author = "Diehl, M.",
    title = "{Generalized parton distributions}",
    eprint = "hep-ph/0307382",
    archivePrefix = "arXiv",
    reportNumber = "DESY-THESIS-2003-018",
    doi = "10.1016/j.physrep.2003.08.002",
    journal = "Phys. Rept.",
    volume = "388",
    pages = "41--277",
    year = "2003"
}

@article{Ji:2004gf,
    author = "Ji, X.",
    title = "{Generalized parton distributions}",
    doi = "10.1146/annurev.nucl.54.070103.181302",
    journal = "Ann. Rev. Nucl. Part. Sci.",
    volume = "54",
    pages = "413--450",
    year = "2004"
}

@article{Belitsky:2005qn,
    author = "Belitsky, A. V. and Radyushkin, A. V.",
    title = "{Unraveling hadron structure with generalized parton distributions}",
    eprint = "hep-ph/0504030",
    archivePrefix = "arXiv",
    reportNumber = "JLAB-THY-04-34",
    doi = "10.1016/j.physrep.2005.06.002",
    journal = "Phys. Rept.",
    volume = "418",
    pages = "1--387",
    year = "2005"
}

@article{Boffi:2007yc,
    author = "Boffi, Sigfrido and Pasquini, Barbara",
    title = "{Generalized parton distributions and the structure of the nucleon}",
    eprint = "0711.2625",
    archivePrefix = "arXiv",
    primaryClass = "hep-ph",
    doi = "10.1393/ncr/i2007-10025-7",
    journal = "Riv. Nuovo Cim.",
    volume = "30",
    number = "9",
    pages = "387--448",
    year = "2007"
}

@article{Diehl:2015uka,
    author = "Diehl, Markus",
    title = "{Introduction to GPDs and TMDs}",
    eprint = "1512.01328",
    archivePrefix = "arXiv",
    primaryClass = "hep-ph",
    reportNumber = "DESY-15-234",
    doi = "10.1140/epja/i2016-16149-3",
    journal = "Eur. Phys. J. A",
    volume = "52",
    number = "6",
    pages = "149",
    year = "2016"
}

@article{Kumericki:2016ehc,
    author = "Kumericki, Kresimir and Liuti, Simonetta and Moutarde, Herve",
    title = "{GPD phenomenology and DVCS fitting}: {Entering the high-precision era}",
    eprint = "1602.02763",
    archivePrefix = "arXiv",
    primaryClass = "hep-ph",
    doi = "10.1140/epja/i2016-16157-3",
    journal = "Eur. Phys. J. A",
    volume = "52",
    number = "6",
    pages = "157",
    year = "2016"
}

@article{Hashamipour:2022noy,
    author = "Hashamipour, Hadi and Goharipour, Muhammad and Azizi, K. and Goloskokov, S. V.",
    title = "{Generalized parton distributions at zero skewness}",
    eprint = "2211.09522",
    archivePrefix = "arXiv",
    primaryClass = "hep-ph",
    doi = "10.1103/PhysRevD.107.096005",
    journal = "Phys. Rev. D",
    volume = "107",
    number = "9",
    pages = "096005",
    year = "2023"
}

@article{Goharipour:2025kif,
    author = "Goharipour, Muhammad and Irani, Fatemeh and Azizi, K. and Dutta, Dipangkar",
    collaboration = "MMGPDs",
    title = "{Fresh look at the nuclear transparency using the generalized parton distributions}",
    eprint = "2507.06333",
    archivePrefix = "arXiv",
    primaryClass = "hep-ph",
    month = "7",
    year = "2025"
}

@article{Bernard:2001rs,
    author = "Bernard, Veronique and Elouadrhiri, Latifa and Meissner, Ulf-G.",
    title = "{Axial structure of the nucleon: Topical Review}",
    eprint = "hep-ph/0107088",
    archivePrefix = "arXiv",
    reportNumber = "FZJ-IKP-TH-01-12, JLAB-PHY-02-53",
    doi = "10.1088/0954-3899/28/1/201",
    journal = "J. Phys. G",
    volume = "28",
    pages = "R1--R35",
    year = "2002"
}

@article{Guidal:2004nd,
    author = "Guidal, M. and Polyakov, M. V. and Radyushkin, A. V. and Vanderhaeghen, M.",
    title = "{Nucleon form-factors from generalized parton distributions}",
    eprint = "hep-ph/0410251",
    archivePrefix = "arXiv",
    reportNumber = "JLAB-THY-04-286",
    doi = "10.1103/PhysRevD.72.054013",
    journal = "Phys. Rev. D",
    volume = "72",
    pages = "054013",
    year = "2005"
}

@article{Diehl:2013xca,
    author = "Diehl, Markus and Kroll, Peter",
    title = "{Nucleon form factors, generalized parton distributions and quark angular momentum}",
    eprint = "1302.4604",
    archivePrefix = "arXiv",
    primaryClass = "hep-ph",
    reportNumber = "DESY-13-025, DESY 13-025",
    doi = "10.1140/epjc/s10052-013-2397-7",
    journal = "Eur. Phys. J. C",
    volume = "73",
    number = "4",
    pages = "2397",
    year = "2013"
}

@article{Polyakov:2018zvc,
    author = "Polyakov, Maxim V. and Schweitzer, Peter",
    title = "{Forces inside hadrons: pressure, surface tension, mechanical radius, and all that}",
    eprint = "1805.06596",
    archivePrefix = "arXiv",
    primaryClass = "hep-ph",
    doi = "10.1142/S0217751X18300259",
    journal = "Int. J. Mod. Phys. A",
    volume = "33",
    number = "26",
    pages = "1830025",
    year = "2018"
}

@article{Adhikari:2021jrh,
    author = "Adhikari, Lekha and Mondal, Chandan and Nair, Sreeraj and Xu, Siqi and Jia, Shaoyang and Zhao, Xingbo and Vary, James P.",
    collaboration = "BLFQ",
    title = "{Generalized parton distributions and spin structures of light mesons from a light-front Hamiltonian approach}",
    eprint = "2110.05048",
    archivePrefix = "arXiv",
    primaryClass = "hep-ph",
    doi = "10.1103/PhysRevD.104.114019",
    journal = "Phys. Rev. D",
    volume = "104",
    number = "11",
    pages = "114019",
    year = "2021"
}

@article{Chavez:2021llq,
    author = "Chavez, Jos{\'e} Manuel Morgado and Bertone, Valerio and De Soto Borrero, Feliciano and Defurne, Maxime and Mezrag, C{\'e}dric and Moutarde, Herv{\'e} and Rodr{\'\i}guez-Quintero, Jos{\'e} and Segovia, Jorge",
    title = "{Pion generalized parton distributions: A path toward phenomenology}",
    eprint = "2110.06052",
    archivePrefix = "arXiv",
    primaryClass = "hep-ph",
    doi = "10.1103/PhysRevD.105.094012",
    journal = "Phys. Rev. D",
    volume = "105",
    number = "9",
    pages = "094012",
    year = "2022"
}

@article{Zhang:2021mtn,
    author = "Zhang, Jin-Li and Raya, Kh{\'e}pani and Chang, Lei and Cui, Zhu-Fang and Morgado, Jos{\'e} Manuel and Roberts, Craig D and Rodr{\'\i}guez-Quintero, Jose",
    title = "{Measures of pion and kaon structure from generalised parton distributions}",
    eprint = "2101.12286",
    archivePrefix = "arXiv",
    primaryClass = "hep-ph",
    reportNumber = "NJU-INP 032/21",
    doi = "10.1016/j.physletb.2021.136158",
    journal = "Phys. Lett. B",
    volume = "815",
    pages = "136158",
    year = "2021"
}

@article{Zhang:2021shm,
    author = "Zhang, Jin-Li and Lai, Meng-Yun and Zong, Hong-Shi and Ping, Jia-Lun",
    title = "{Pion generalized parton distributions and light-front wave functions in the Nambu{\textendash}Jona-Lasinio model}",
    doi = "10.1016/j.nuclphysb.2021.115387",
    journal = "Nucl. Phys. B",
    volume = "966",
    pages = "115387",
    year = "2021"
}

@article{Raya:2021zrz,
    author = "Raya, Khepani and Cui, Zhu-Fang and Chang, Lei and Morgado, Jose-Manuel and Roberts, Craig D. and Rodriguez-Quintero, Jose",
    title = "{Revealing pion and kaon structure via generalised parton distributions *}",
    eprint = "2109.11686",
    archivePrefix = "arXiv",
    primaryClass = "hep-ph",
    reportNumber = "NJU-INP 051/21",
    doi = "10.1088/1674-1137/ac3071",
    journal = "Chin. Phys. C",
    volume = "46",
    number = "1",
    pages = "013105",
    year = "2022"
}

@article{Cao:2021aci,
    author = "Cao, N. Y. and Barry, P. C. and Sato, N. and Melnitchouk, W.",
    collaboration = "Jefferson Lab Angular Momentum",
    title = "{Towards the three-dimensional parton structure of the pion: Integrating transverse momentum data into global QCD analysis}",
    eprint = "2103.02159",
    archivePrefix = "arXiv",
    primaryClass = "hep-ph",
    reportNumber = "JLAB-THY-21-3328",
    doi = "10.1103/PhysRevD.103.114014",
    journal = "Phys. Rev. D",
    volume = "103",
    number = "11",
    pages = "114014",
    year = "2021"
}

@article{Broniowski:2022iip,
    author = "Broniowski, Wojciech and Shastry, Vanamali and Ruiz Arriola, Enrique",
    title = "{Off-shell generalized parton distributions and form factors of the pion}",
    eprint = "2211.11067",
    archivePrefix = "arXiv",
    primaryClass = "hep-ph",
    doi = "10.1016/j.physletb.2023.137872",
    journal = "Phys. Lett. B",
    volume = "840",
    pages = "137872",
    year = "2023"
}

@article{Lin:2023gxz,
    author = "Lin, Huey-Wen",
    title = "{Pion valence-quark generalized parton distribution at physical pion mass}",
    eprint = "2310.10579",
    archivePrefix = "arXiv",
    primaryClass = "hep-lat",
    reportNumber = "MSUHEP-23-011",
    doi = "10.1016/j.physletb.2023.138181",
    journal = "Phys. Lett. B",
    volume = "846",
    pages = "138181",
    year = "2023"
}

@article{Xu:2023bwv,
    author = "Xu, Yin-Zhen and Raya, Kh{\'e}pani and Cui, Zhu-Fang and Roberts, Craig D. and Rodr{\'\i}guez-Quintero, J.",
    title = "{Empirical Determination of the Pion Mass Distribution}",
    eprint = "2302.07361",
    archivePrefix = "arXiv",
    primaryClass = "hep-ph",
    reportNumber = "NJU-INP 070/23",
    doi = "10.1088/0256-307X/40/4/041201",
    journal = "Chin. Phys. Lett.",
    volume = "40",
    number = "4",
    pages = "041201",
    year = "2023"
}

@article{Ding:2024saz,
    author = "Ding, Heng-Tong and Gao, Xiang and Mukherjee, Swagato and Petreczky, Peter and Shi, Qi and Syritsyn, Sergey and Zhao, Yong",
    title = "{Three-dimensional imaging of pion using lattice QCD: generalized parton distributions}",
    eprint = "2407.03516",
    archivePrefix = "arXiv",
    primaryClass = "hep-lat",
    doi = "10.1007/JHEP02(2025)056",
    journal = "JHEP",
    volume = "02",
    pages = "056",
    year = "2025"
}

@article{Son:2024uet,
    author = "Son, Hyeon-Dong and Hutauruk, Parada T. P.",
    title = "{Generalized parton distributions of the kaon and pion within the nonlocal chiral quark model}",
    eprint = "2411.18130",
    archivePrefix = "arXiv",
    primaryClass = "hep-ph",
    doi = "10.1103/PhysRevD.111.054007",
    journal = "Phys. Rev. D",
    volume = "111",
    number = "5",
    pages = "054007",
    year = "2025"
}

@article{Nematollahi:2024wrj,
    author = "Nematollahi, H. and Azizi, K.",
    title = "{Unpolarized valence GPDs and form factors of the pion in the modified chiral quark model}",
    eprint = "2412.13322",
    archivePrefix = "arXiv",
    primaryClass = "hep-ph",
    doi = "10.1103/PhysRevD.111.014011",
    journal = "Phys. Rev. D",
    volume = "111",
    number = "1",
    pages = "014011",
    year = "2025"
}

@article{Puhan:2025ibn,
    author = "Puhan, Satyajit and Kaur, Navpreet and Kumar, Arvind and Dutt, Suneel and Dahiya, Harleen",
    title = "{Effect of nuclear medium on the spatial distribution of pions}",
    eprint = "2501.16706",
    archivePrefix = "arXiv",
    primaryClass = "hep-ph",
    doi = "10.1016/j.nuclphysb.2025.116940",
    journal = "Nucl. Phys. B",
    volume = "1017",
    pages = "116940",
    year = "2025"
}

@article{Nasibova:2025wnw,
    author = "Nasibova, Narmin and Arsiwalla, Xerxes D.",
    title = "{Pion Phenomenology from the Thermal Soft-Wall Model of Holographic QCD}",
    eprint = "2505.23455",
    archivePrefix = "arXiv",
    primaryClass = "hep-ph",
    month = "5",
    year = "2025"
}

@article{Zhang:2025xtn,
    author = "Zhang, Jin-Li",
    title = "{Off-shell modifications of the pion generalized parton distributions and transverse momentum dependent parton distributions}",
    eprint = "2507.09557",
    archivePrefix = "arXiv",
    primaryClass = "hep-ph",
    month = "7",
    year = "2025"
}

@article{Lepage:1980fj,
    author = "Lepage, G. Peter and Brodsky, Stanley J.",
    title = "{Exclusive Processes in Perturbative Quantum Chromodynamics}",
    reportNumber = "SLAC-PUB-2478",
    doi = "10.1103/PhysRevD.22.2157",
    journal = "Phys. Rev. D",
    volume = "22",
    pages = "2157",
    year = "1980"
}

@article{Bakulev:2004cu,
    author = "Bakulev, A. P. and Passek-Kumericki, K. and Schroers, W. and Stefanis, N. G.",
    title = "{Pion form-factor in QCD: From nonlocal condensates to NLO analytic perturbation theory}",
    eprint = "hep-ph/0405062",
    archivePrefix = "arXiv",
    reportNumber = "IRB-TH-02-04, MIT-CTP-3478, RUB-TPII-02-04",
    doi = "10.1103/PhysRevD.70.033014",
    journal = "Phys. Rev. D",
    volume = "70",
    pages = "033014",
    year = "2004",
    note = "[Erratum: Phys.Rev.D 70, 079906 (2004)]"
}

@article{Chen:2023byr,
    author = "Chen, Long-Bin and Chen, Wen and Feng, Feng and Jia, Yu",
    title = "{Next-to-Next-to-Leading-Order QCD Corrections to Pion Electromagnetic Form Factors}",
    eprint = "2312.17228",
    archivePrefix = "arXiv",
    primaryClass = "hep-ph",
    doi = "10.1103/PhysRevLett.132.201901",
    journal = "Phys. Rev. Lett.",
    volume = "132",
    number = "20",
    pages = "201901",
    year = "2024",
    note = "[Erratum: Phys.Rev.Lett. 134, 229901 (2025)]"
}

@article{Wang:2025irh,
    author = "Wang, Sheng-Quan and Liao, Zuo-Fen and Shen, Jian-Ming and Zhou, Hua and Zhang, Jia-Wei and Yan, Jiang and Wu, Xing-Gang and Di Giustino, Leonardo",
    title = "{Analysis of the Pion Electromagnetic Form Factor with Next-to-Next-to-Leading Order QCD Corrections}",
    eprint = "2507.20479",
    archivePrefix = "arXiv",
    primaryClass = "hep-ph",
    month = "7",
    year = "2025"
}

@article{Boyle:2008yd,
    author = "Boyle, P. A. and Flynn, J. M. and Juttner, A. and Kelly, C. and de Lima, H. Pedroso and Maynard, C. M. and Sachrajda, C. T. and Zanotti, J. M.",
    title = "{The Pion's electromagnetic form-factor at small momentum transfer in full lattice QCD}",
    eprint = "0804.3971",
    archivePrefix = "arXiv",
    primaryClass = "hep-lat",
    reportNumber = "EDINBURGH-2008-18, MKPH-T-08-07, SHEP-08-15",
    doi = "10.1088/1126-6708/2008/07/112",
    journal = "JHEP",
    volume = "07",
    pages = "112",
    year = "2008"
}

@article{Wang:2020nbf,
    author = "Wang, Gen and Liang, Jian and Draper, Terrence and Liu, Keh-Fei and Yang, Yi-Bo",
    collaboration = "chiQCD",
    title = "{Lattice Calculation of Pion Form Factor with Overlap Fermions}",
    eprint = "2006.05431",
    archivePrefix = "arXiv",
    primaryClass = "hep-ph",
    doi = "10.1103/PhysRevD.104.074502",
    journal = "Phys. Rev. D",
    volume = "104",
    pages = "074502",
    year = "2021"
}

@article{Ding:2024lfj,
    author = "Ding, Heng-Tong and Gao, Xiang and Hanlon, Andrew D. and Mukherjee, Swagato and Petreczky, Peter and Shi, Qi and Syritsyn, Sergey and Zhang, Rui and Zhao, Yong",
    title = "{QCD Predictions for Meson Electromagnetic Form Factors at High Momenta: Testing Factorization in Exclusive Processes}",
    eprint = "2404.04412",
    archivePrefix = "arXiv",
    primaryClass = "hep-lat",
    doi = "10.1103/PhysRevLett.133.181902",
    journal = "Phys. Rev. Lett.",
    volume = "133",
    number = "18",
    pages = "181902",
    year = "2024"
}

@article{Gao:2021xsm,
    author = "Gao, Xiang and Karthik, Nikhil and Mukherjee, Swagato and Petreczky, Peter and Syritsyn, Sergey and Zhao, Yong",
    title = "{Pion form factor and charge radius from lattice QCD at the physical point}",
    eprint = "2102.06047",
    archivePrefix = "arXiv",
    primaryClass = "hep-lat",
    doi = "10.1103/PhysRevD.104.114515",
    journal = "Phys. Rev. D",
    volume = "104",
    number = "11",
    pages = "114515",
    year = "2021"
}

@article{Chang:2013nia,
    author = {Chang, L. and Clo{\"e}t, I. C. and Roberts, C. D. and Schmidt, S. M. and Tandy, P. C.},
    title = "{Pion electromagnetic form factor at spacelike momenta}",
    eprint = "1307.0026",
    archivePrefix = "arXiv",
    primaryClass = "nucl-th",
    doi = "10.1103/PhysRevLett.111.141802",
    journal = "Phys. Rev. Lett.",
    volume = "111",
    number = "14",
    pages = "141802",
    year = "2013"
}

@article{Jia:2024dfl,
    author = {Jia, Shaoyang and Clo{\"e}t, Ian},
    title = "{Pion Electromagnetic Form Factor from Bethe-Salpeter Amplitudes with Appropriate Kinematics}",
    eprint = "2402.00285",
    archivePrefix = "arXiv",
    primaryClass = "hep-ph",
    month = "1",
    year = "2024"
}

@article{Brodsky:2007hb,
    author = "Brodsky, Stanley J. and de Teramond, Guy F.",
    title = "{Light-Front Dynamics and AdS/QCD Correspondence: The Pion Form Factor in the Space- and Time-Like Regions}",
    eprint = "0707.3859",
    archivePrefix = "arXiv",
    primaryClass = "hep-ph",
    reportNumber = "SLAC-PUB-12554",
    doi = "10.1103/PhysRevD.77.056007",
    journal = "Phys. Rev. D",
    volume = "77",
    pages = "056007",
    year = "2008"
}

@article{Simula:2023ujs,
    author = "Simula, Silvano and Vittorio, Ludovico",
    title = "{Dispersive analysis of the experimental data on the electromagnetic form factor of charged pions at spacelike momenta}",
    eprint = "2309.02135",
    archivePrefix = "arXiv",
    primaryClass = "hep-ph",
    doi = "10.1103/PhysRevD.108.094013",
    journal = "Phys. Rev. D",
    volume = "108",
    number = "9",
    pages = "094013",
    year = "2023"
}

@article{Nesterenko:1982gc,
    author = "Nesterenko, V. A. and Radyushkin, A. V.",
    title = "{Sum Rules and Pion Form-Factor in QCD}",
    reportNumber = "JINR-E2-82-204",
    doi = "10.1016/0370-2693(82)90528-7",
    journal = "Phys. Lett. B",
    volume = "115",
    pages = "410",
    year = "1982"
}

@article{Ayala:2025wuu,
    author = "Ayala, Cesar and Mikhailov, S. V. and Pimikov, A. V.",
    title = "{Extending the LCSR method to the electromagnetic pion form factor at low momenta using QCD renormalization-group summation}",
    eprint = "2505.18349",
    archivePrefix = "arXiv",
    primaryClass = "hep-ph",
    month = "5",
    year = "2025"
}

@article{Xu:2023izo,
    author = "Xu, Yin-Zhen and Ding, Minghui and Raya, Kh{\'e}pani and Roberts, Craig D. and Rodr{\'\i}guez-Quintero, Jos{\'e} and Schmidt, Sebastian M.",
    title = "{Pion and kaon electromagnetic and gravitational form factors}",
    eprint = "2311.14832",
    archivePrefix = "arXiv",
    primaryClass = "hep-ph",
    reportNumber = "NJU-INP 081/23",
    doi = "10.1140/epjc/s10052-024-12518-x",
    journal = "Eur. Phys. J. C",
    volume = "84",
    number = "2",
    pages = "191",
    year = "2024"
}

\end{document}